\documentclass{ws-ijmpd}

\usepackage{graphicx}  
\usepackage{latexsym}  

\def\rmd{\rm d}

\begin{document}

%

\def\nocropmarks{\vskip5pt\phantom{cropmarks}}

\let\trimmarks\nocropmarks      

%

\markboth{Bini D., Geralico A.}
{Superposition of Weyl solutions: circular orbits
}

%
\catchline{}{}{}
%

\title{SUPERPOSITION OF WEYL SOLUTIONS: CIRCULAR ORBITS
}

\author{\footnotesize DONATO BINI}

\address{Istituto per le Applicazioni del Calcolo ``M. Picone'', CNR I-00161 Rome, Italy  \\
International Center for Relativistic Astrophysics - I.C.R.A.,
University of Rome ``La Sapienza'', I-00185 Rome, Italy
\footnote{binid@icra.it}
}

\author{ANDREA GERALICO}

\address{
Dipartimento di Fisica, Universit\`a di Lecce, and INFN, Sezione di Lecce, Via Arnesano, CP 193, I-73100 Lecce, Italy\\
International Center for Relativistic Astrophysics - I.C.R.A.,
University of Rome ``La Sapienza'', I-00185 Rome, Italy
\footnote{geralico@icra.it}
}

\maketitle

\begin{history}
\received{}
\revised{}
\end{history}

\begin{abstract}
Circular orbits are examined in static spacetimes belonging to the Weyl class of vacuum solutions  which represent (nonlinear) superposition of the gravitational fields generated by certain collinear distributions of matter. 
In particular, solutions representing  two and three Chazy-Curzon particles - all of them endowed with conical singularities - are considered. Conditions for geodesic motion in certain symmetry planes are discussed and results are summarized in a number of graphics too. All the discussion is developed in the framework of observer-dependent analysis of motion.
\end{abstract}


\section{Introduction}

Axisymmetric, static, vacuum solutions of the Einstein field equations can be described by the Weyl formalism \cite{weyl}. The corresponding line element in Weyl  canonical coordinates \cite{exactsols} $(t,\rho,z,\phi)$ is 
\begin{equation}
\label{weylmetric}
ds^2=-e^{2\psi}dt^2+e^{2(\gamma-\psi)}[d\rho^2+dz^2]+\rho^2e^{-2\psi}d\phi^2\ ,
\end{equation}
where the function $\psi$ and $\gamma$ depend on coordinates $\rho$ and $z$ only. The vacuum Einstein field equations reduce to the following 
\begin{eqnarray}
\label{einsteqs}
&& \qquad \qquad \psi_{,\rho\rho}+\frac1\rho\psi_{,\rho}+\psi_{,zz}=0, \nonumber\\
&& \gamma_{,\rho}-\rho[\psi_{,\rho}^2-\psi_{,z}^2]=0, \qquad
\gamma_{,z}-2\rho\psi_{,\rho}\psi_{,z}=0\ .
\end{eqnarray}
The linearity of the first equation of (\ref{einsteqs}) allows to find solutions which represent superpositions of two or more axially symmetric bodies of equal or different shapes explicitly \cite{letelier}. In general such a configuration is not gravitationally stable. Indeed, the
nonlinear terms of the remaining equations are responsible for the occurrence of gravitationally inert singular
structures, ``struts'' and ``membranes'', that keep the bodies apart making the configuration as stable.
In this paper we are concerned in analyzing circular orbits in spacetimes representing superpositions of two or three Chazy-Curzon particles. However, the whole discussion (and results) can be easily extended
to the analogous situation of superpositions of two or three Schwarzschild black holes, or in the mixed case of Schwarzschild black hole and Chazy-Curzon particle fields superposed, as (briefly) shown in the Appendix. 

It can be shown (see, e.g., \cite{sokolov}) that all these solutions are characterized by a conical singularity on the $z-$axis, the occurrence of which is related to the non-vanishing of the function $\gamma(\rho,z)$ on the portion of the axis between the sources or outside them. In the former case we can interpret the singular segment of the axis as a strut holding the bodies apart, while in the latter one as a pair of cords on which the bodies are suspended. The strut is represented by a perfect fluid with a pressure and a negative energy density equal to it; qualitatively, the energy of the strut can be thought as the interaction energy between the bodies, while its pressure prevents the gravitational collapse of the system. The effective gravitational mass of the strut vanishes \cite{israel}; as a consequence, it makes no contribution to the gravitational potential $\psi(\rho,z)$. 

To our knowledge, the discussion present in the literature on this topic is very limited, a part for the pioneering works of Semer\'ak, Zellerin and ${\check {\rm Z}}$\'a${\check {\rm c}}$ek \cite{sem1,sem2}, who were but interested in solutions representing stationary axisymmetric sources around black holes, i.e.  accretion disks or rings of astrophysical interest. The reason for which this class of solutions is poorly studied is mainly due to the presence of singularities, often dismissed as representative of non-physical situations.
On the other hand, singularities are somehow typical in general relativity and it is especially in order to better understand their role and their proper character in this quite simple class of solutions that  the present paper has been conceived. 
Circular orbits are very well known for general stationary (and static) spacetimes \cite{iyevis,def95,pag98,sem98,bjdf0,circfs,bjdf,mfg,idcf1,idcf2} and they are considered here as a convenient tool to perform the above mentioned analysis. The results will be mainly discussed by graphics because of the very long formulas involved in the treatment of such solutions.

\section{Circular orbits}

Let us consider a family of test particles moving along the $\phi$ direction with constant speed; the (timelike) 4-velocity $U$ associated to a generic orbit within the family is the following:
\begin{equation}
U=\Gamma_{\zeta}[\partial_t + \zeta\partial_{\phi}]=\gamma_n [n +\nu  e_{\hat\phi}]=\cosh \alpha\,  n +\sinh \alpha\,  e_{\hat\phi}\ ,
\end{equation}
where $\zeta$, $\nu$ or $\alpha$ are the angular velocity, linear velocity and rapidity parametrization respectively of the whole family, satisfying the mutual relations
\begin{equation}
\nu=e^{-2\psi}\,\rho \zeta=\tanh \alpha.
\end{equation}
$\Gamma_{\zeta}$ is defined by the timelike condition $U\cdot U=-1$ as
\begin{equation}
-\Gamma_{\zeta}^{-2}=g_{tt}+\zeta^2g_{\phi\phi}=-e^{2\psi}+\zeta^2\rho^2e^{-2\psi}=-\frac{e^{2\psi}}{\gamma_n^2}\ ;
\end{equation}
$n=e^{-\psi}\partial_t$ is the four-velocity of the standard family of static observers with 
$e_{\hat \rho}=e^{\psi-\gamma}\partial_\rho$, $e_{\hat z}=e^{\psi-\gamma}\partial_z$, and $e_{\hat \phi}=\frac{e^\psi}{\rho}\partial_\phi$
being the associated orthonormal spatial directions; finally $\gamma_n=-U\cdot n=(1-\nu^2)^{-1/2}=\Gamma_\zeta e^\psi$ is the Lorentz factor.

Even if we are concerned here with timelike orbits, it is worth noticing that the case of null orbits for co-rotating and counter-rotating photons correspond to the angular velocities:
\begin{equation}
\zeta_{{\rm null}\ \pm}=\pm \frac{e^{2\psi}}{\rho}, \qquad \nu_{{\rm null}\ \pm}=\pm 1.
\end{equation}

The non-vanishing components of the 4-acceleration $a(U)=\nabla_U U$ along $U$ are given by
\begin{eqnarray}
\label{acccomp}
a(U)^{\hat \rho}&=&\frac{e^{\psi-\gamma}}{e^{4\psi}-\rho^2\zeta^2}\left[
\psi_{,\rho} (e^{4\psi}+\rho^2\zeta^2)-\rho\zeta^2\right]=e^{\psi-\gamma} \gamma_n^2 [\psi_{,\rho} -\frac{\nu^2}{\rho}(1-\rho \psi_{,\rho})] ,\nonumber \\
a(U)^{\hat z}&=&e^{\psi-\gamma}\psi_{,z}\frac{e^{4\psi}+\rho^2\zeta^2}{e^{4\psi}-\rho^2\zeta^2}= e^{\psi-\gamma}\psi_{,z}\gamma_n^2 (1+\nu^2) \ .
\end{eqnarray}
The lines of force of the acceleration or \lq\lq gravitoelectric'' field associated to a single orbit (i.e. with fixed $\nu$) are defined by the condition
\begin{equation}
\label{linesofforce}
\frac{\rmd \rho}{{\rm d} z}=\frac{a^{\hat \rho}}{a^{\hat z}}=\frac{\psi_{,\rho}}{\psi_{,z}}-\frac{1}{\rho \psi_{,z}}\, \frac{\nu^2}{1+\nu^2}.
\end{equation}

In the same way as one can consider the family of all the circular orbits as describing a branch of hyperbola in the relative motion $t - \phi$ plane,   equations (\ref{acccomp}) are the parametric equations for an \lq\lq acceleration curve'' in the $\{ a^{\hat \rho},a^{\hat z}\}$ acceleration $\rho - z$ plane; eliminating the parameter $\nu$, the cartesian equation of this acceleration curve here (at $\rho$ and $z$ fixed) is simply a straight line
\begin{equation}
a(U)^{\hat z}=- \frac{2\rho \psi_{,z}}{1-2\rho \psi_{,\rho}}\, a(U)^{\hat \rho}+ \frac{e^{\psi-\gamma} \psi_{,z}}{1-2\rho \psi_{,\rho}} .
\end{equation}
The absolute value of the acceleration is
\begin{equation}
\kappa=||a(U)||=e^{\psi-\gamma} \gamma_n^2 \left[ \left(\psi_{,\rho} -\frac{\nu^2}{\rho}(1-\rho \psi_{,\rho})\right)^2
+\psi_{,z}^2 (1+\nu^2)^2 \right]^{1/2}\ ,
\end{equation}
and it is symmetric as a function of $\nu$;
one can also introduce polar coordinates in the acceleration plane: $(\kappa, \chi)$ so that
\begin{equation}
a(U)^{\hat \rho}=\kappa \cos\chi\ , \qquad 
a(U)^{\hat z}=\kappa \sin \chi\ , \qquad 
\tan \chi = \frac{\psi_{,z}(1+\nu^2)}{\psi_{,\rho} -\frac{\nu^2}{\rho}(1-\rho \psi_{,\rho})}\ .
\end{equation}
Moreover, the Lie relative curvature \cite{idcf1,idcf2} of the orbits is $k_{\rm (lie)}= -\nabla\ln{(\sqrt{g_{\phi\phi}})}$, with components
\begin{equation}
k_{\rm (lie)}{}_{\hat \rho} = - e^{\psi-\gamma}\frac{1-\rho \psi_{,\rho}}{\rho}\ . 
\qquad  k_{\rm (lie)}{}_{\hat z} = e^{\psi-\gamma}\psi_{,z}\ ,
\end{equation}
One can also introduce a polar representation for $k_{\rm (lie)}$, 
\begin{equation}
k_{\rm (lie)}{}_{\hat \rho}=\kappa_{\rm (lie)}\cos \chi_{\rm (lie)}, \quad k_{\rm (lie)}{}_{\hat z}=\kappa_{\rm (lie)}\sin \chi_{\rm (lie)},\quad
\tan \chi_{\rm (lie)}= \frac{\rho \psi_{,z}}{\rho \psi_{,\rho}-1} 
\end{equation}
and study the corresponding lines of force defined by
\begin{equation}
\label{lfklie}
\frac{{\rm d} \rho}{{\rm d}z}=\frac{g_{\phi\phi ,\rho}}{g_{\phi\phi ,z}}\ , 
\end{equation}
showing also that this vector is orthogonal to the $g_{\phi\phi}=const$ hypersurfaces.

The discussion presented above is quite standard now and it follows the notation of \cite{bjdf}. 
Special orbits can be selected so that
\begin{equation}
\zeta_{\pm}=\pm e^{2\psi}\left[\frac{\psi_{,\rho}}{\rho(1-\rho\psi_{,\rho})}\right]^{1/2}, \qquad
\nu_{\pm}=\pm \left[-1+\frac{1}{\rho\psi_{,\rho}}\right]^{-1/2}, 
\end{equation}
which permits the component $a^{\hat \rho}$ to vanish; the quantities $\Gamma_{\zeta}$ and $a^{\hat z}$ then become
\begin{eqnarray}
\Gamma_{\zeta_{\pm}}=e^{-\psi}\left[\frac{1-\rho\psi_{,\rho}}{1-2\rho\psi_{,\rho}}\right]^{1/2}\ , \qquad
a^{\hat z}\big\vert_{\zeta=\zeta_{\pm}}=e^{\psi-\gamma}\frac{\psi_{,z}}{1-2\rho\psi_{,\rho}} \ .
\end{eqnarray}
The further requirement $a^{\hat z}=0$ (and so $\psi_{,z}=0$) gives the conditions for circular geodesics expressed in terms of the parameters of the sources and their relative position.
These geodesics become null at the radius such that $\psi_{,\rho}=1/2\rho$.
Moreover, orbits corresponding to extremal values of $\kappa$ and $\chi$ among the whole family are also special; 
for instance, the condition 
\begin{equation}
\frac{\rmd \kappa}{\rmd \nu}=0, 
\end{equation}
selects the \lq\lq extremely accelerated'' orbits
\begin{equation}
\nu_{(\kappa,\rm  ext)}{}_{(0)}=0\ , \nu_{(\kappa,\rm  ext)}{}_{(\pm)}=\pm \left[\frac{1-2\rho\psi_{,\rho}}{2\rho^2(\psi_{,\rho}^2+\psi_{,z}^2)-3\rho\psi_{,\rho}+1}-1\right]^{1/2}\ ,
\end{equation}
and it can be interpreted in the ambit of the Frenet-Serret formalism \cite{bjdf} giving the result that the orbits have vanishing the Frenet-Serret first torsion.
The complementary condition
\begin{equation}
\frac{\rmd \chi}{\rmd \nu}=0, 
\end{equation}
gives only $\nu_{(\chi , {\rm ext)}}=0$,
which can  also be interpreted in the ambit of the Frenet-Serret formalism \cite{bjdf} giving the result that the orbits have vanishing the Frenet-Serret second torsion.
It is easy to show that $\nu_{(\kappa,\rm  ext)}{}_{(\pm)}=\pm 1$ at the same radius at which the geodesics become null: $\psi_{,z}=0$, $\nu_\pm =\pm 1$.
Finally, by using $\nu_\pm$ and $k_{\rm (lie)}$ the components of the acceleration can be cast in the form
\begin{equation}
a(U)_{\hat \rho}=k_{\rm (lie)}{}_{\hat \rho}\, \gamma_n^2 (\nu^2-\nu_\pm^2), \qquad 
a(U)_{\hat z}= k_{\rm (lie)}{}_{\hat z} \, \gamma_n^2 (1+\nu^2),
\end{equation}
and the relation between $\chi$ and $\chi_{\rm (lie)}$ becomes
\begin{equation}
\tan \chi = \frac{1+\nu^2}{\nu^2-\nu^2_\pm}\, \tan \chi_{\rm (lie)} . 
\end{equation}

\section{Solutions representing two or three Chazy-Curzon particles}

In this section we discuss the cases of superimposed Weyl fields due to two or three Chazy-Curzon particles.

A single Chazy-Curzon particle is a static axisymmetric solution of Einstein equations endowed with a singularity at the particle position \cite{chazy,curzon}. 
The Curzon metric is generated by the newtonian potential of a spherically symmetric point mass using the Weyl formalism (see (\ref{ccsolw})), and can be obtained from 
the correspondent potential representing a Schwarzschild black hole (whose generating newtonian potential (\ref{SSbsol}) is, instead, that of a line mass) by linearizing it with respect to the mass of the rod.
The singularity is not point-like, but rather ring-like \cite{scott}, exhibiting directional properties. In fact, at the particle location the first Kretschmann invariant ${\mathcal K}=R^{\alpha \beta \gamma \delta }R_{\alpha \beta \gamma \delta }$
can be zero, infinite or finite, according to the direction of approach to the singularity. For instance, approaching the singularity along the $z-$axis the Kretschmann invariant ${\mathcal K}\rightarrow 0$; as a consequence
all the geodesics which approach the particle location, either along the $z-$axis or tending to it asymptotically, do not meet a curvature singularity in the sense of ${\mathcal K}$ diverging at that point.

We first consider the case of one particle located at the  origin 
$(\rho , z)=(0,0)$ and  the other displaced along the $z-$axis  at the position $(\rho , z)=(0,b)$, with $b>0$  satisfying a certain condition expressed in terms of the particle or black hole parameters; then we study the addition of a third particle at the position  $(\rho , z)=(0,c)$ with $c>b$.
The properties of the circular orbits at any radius and at different values of the coordinate $z$ and varying the distance parameter $b$ are studied with the aid of graphics.

In the Appendix we put details necessary to repeat the whole discussion presented here in the case in which the Weyl fields are due to two Schwarzschild black holes and to a combination of both Schwarzschild black holes and Chazy-Curzon particles. Extensions to cases involving any number of such particles can be easily treated too.

\subsection{Superposition of two Chazy-Curzon particles}

The solution corresponding to the superposition of two Chazy-Curzon particles with masses ${\mathcal M}$ and $m$ and positions $z=0$ and $z=b$ on the $z-$axis respectively is given by metric (\ref{weylmetric}) with functions
\begin{eqnarray}
\label{psigammaCCb}
\psi=\psi_{\rm C}+\psi_{\rm C_b}\ , \qquad \gamma=\gamma_{\rm C}+\gamma_{\rm C_b}+\gamma_{\rm CC}\ ,
\end{eqnarray}
where
\begin{eqnarray}
\label{ccsolw}
\psi_{\rm C}&=&-\frac{{\mathcal M}}{R_{\rm C}}\ , \qquad \gamma_{\rm C}=-\frac12\frac{{\mathcal M}^2\rho^2}{R_{\rm C}^4}\ , \qquad R_{\rm C}=\sqrt{\rho^2+z^2}\ ,  \nonumber\\
\psi_{\rm C_b}&=&-\frac{m}{R_{\rm C_b}}\ , \qquad \gamma_{\rm C_b}=-\frac12\frac{m^2\rho^2}{R_{\rm C_b}^4}\ , \qquad R_{\rm C_b}=\sqrt{\rho^2+(z-b)^2}\ 
\end{eqnarray}
and $\gamma_{\rm CC}$ can be obtained by solving Einstein's equations (\ref{einsteqs}):
\begin{equation}
\gamma_{\rm CC}=2\frac{m{\mathcal M}}{b^2}\frac{\rho^2+z(z-b)}{R_{\rm C_b}R_{\rm C}}+C\ .
\end{equation}
The value of arbitrary constant $C$ can be determined by imposing the elementary flatness condition
\begin{equation}
\label{regcond}
\lim_{\rho\rightarrow0}\gamma=0\ ,
\end{equation}
from which we have
\begin{equation}
0=\lim_{\rho\rightarrow0}\gamma_{\rm CC}=2\frac{m{\mathcal M}}{b^2}\frac{z(z-b)}{|z||z-b|}+C\ .
\end{equation}
The relevant cases to analyze are the following:
\begin{itemize}
\item{case 1: $z>b$ (and, analogously, $z<0$):
\begin{equation}
0=\gamma_{\rm CC}(0,z)=2\frac{m{\mathcal M}}{b^2}+C\ ;
\end{equation}
}
\item{case 2: $0<z<b$:
\begin{equation}
0=\gamma_{\rm CC}(0,z)=-2\frac{m{\mathcal M}}{b^2}+C\ .
\end{equation}
}
\end{itemize}
The arbitrary constant $C$ cannot be uniquely chosen in order the function $\gamma_{\rm CC}$ to vanish on the whole $z-$axis: a $\gamma_{\rm CC}\not= 0$ gives rise to the well known conical singularity, corresponding to a strut in compresson, which holds the two particles apart. 
The choice $C=2m{\mathcal M}/b^2$ makes $\gamma_{\rm CC}=0$ only on the segment $0<z<b$ of the $z-$axis between the sources. 
In the following we use $C=-2m{\mathcal M}/b^2$, that makes $\gamma_{\rm CC}=0$ on the portion of the axis outside the sources ($z<0$ and $z>b$).

In Section 2 we have shown that the condition $\kappa=0$ for geodesic motion yields to the requirements $\psi_{,z}=0$ and $\nu=\nu_\pm$, with $|\nu_\pm|<1$, from which 
\begin{eqnarray}
\label{CCgeos}
\frac{{\mathcal M}z}{R_{\rm C}^3}+\frac{m(z-b)}{R_{\rm C_b}^3}=0\ , \qquad
\left[-1+\frac1{\rho^2}\left(\frac{{\mathcal M}}{R_{\rm C}^3}+\frac{m}{R_{\rm C_b}^3}\right)^{-1}\right]^{-1/2}<1\ 
\end{eqnarray}
respectively. Eq. (\ref{CCgeos})$_1$ admits solutions only in the region between the sources, namely for $0<z<b$, and only for certain selected values of $\rho$ (the roots of the equation), once the parameters $b$, ${\mathcal M}$ and $m$ are fixed. 
However, there exists a special solution for equal masses ${\mathcal M}=m$ on the middle plane $z=b/2$, for which the equation is satisfied for all $\rho$, whose range of allowed values is defined by the condition (\ref{CCgeos})$_2$, that writes in this case as follows:
\begin{equation}
\label{rhoeq}
\left[-1+\frac1{2m\rho^2}\left[\rho^2+\frac{b^2}4\right]^{3/2}\right]^{-1/2}<1\ .
\end{equation}
By introducing the new variable $\xi$, $\rho ^2=-b^2/4+\xi ^2$ and squaring both sides eq. (\ref{rhoeq}) becomes,
\begin{equation}
\label{etaCCgeoscond}
\xi^3-4m\xi^2+b^2>0\ , \qquad \xi\in\left[b/2, \infty\right)\ ,
\end{equation}
and it can be cast in the standard form
\begin{equation}
\label{yCCgeoscond}
\eta^3+p\eta+q>0\ , \qquad \eta\in\left[\eta_0, \infty\right)\ , \qquad \eta _0=\frac{b}2-\frac43m\ ,
\end{equation}
by making the substitution $\eta=\xi-4m/3$, with
\begin{equation}
p=-\frac{16}3m^2\ , \qquad q=\frac{16}{27}m\left[\frac{27}{16}b^2-8m^2\right]\ .
\end{equation}
The solution of equation (\ref{yCCgeoscond}) can be written as
\begin{equation}
\label{cardano}
\eta_1={\mathcal A}_{+}+{\mathcal A}_{-}\ , \qquad \eta_2=\frac12\left[\eta _1+i\sqrt{3}({\mathcal A}_{+}-{\mathcal A}_{-})\right]\ , \qquad \eta_3={\bar \eta_2}\ ,
\end{equation}
(the bar denotes the operation of complex conjugation) where the quantities ${\mathcal A}_\pm$ and $\Delta$ are given by
\begin{equation}
{\mathcal A}_\pm=\left[-\frac{q}2\pm\sqrt{\Delta}\right]^{1/3}\ , \qquad \Delta=\frac{q^2}4+\frac{p^3}{27}\ .
\end{equation}
The character of the solution depends on the sign of the polynomial discriminant $\Delta $: if $\Delta>0$ one root is real, and the others two are complex conjugates; 
if $\Delta\leq0$ all the roots are real, at least two of them coinciding if $\Delta=0$. 
The real solutions (three unequal roots, which correspond to have $\Delta<0$) are of the form:
\begin{equation}
\eta_k=2\sqrt{-\frac{p}3}\cos{\frac{\delta+2k\pi}3}\ , \qquad k=0,1,2\ , 
\end{equation}
where $\delta=\omega$ if $-q/2>0$, or $\delta=\pi+\omega$ if $-q/2<0$, being $\omega=\arctan{(-2\sqrt{-\Delta}/q)}$.
The requirement $\Delta<0$ gives the following condition on the distance parameter $b$:
\begin{equation}
\frac{27}{16}b^2-16m^2<0\ , 
\end{equation}
so that $b\in[0, {\tilde b})$, ${\tilde b}=16\sqrt3 m/9$.
In terms of the initial variable $\rho$ the solutions of eq. (\ref{rhoeq}) are given by (the hat indicates that they are ordered)
\begin{equation}
{\hat \rho}_i=\left[\left({\hat \eta}_i+\frac43m\right)^2-\frac{b^2}4\right]^{1/2}\ , \qquad i=1,2,3\ . 
\end{equation}
The analysis of the roots leads to the conclusion that timelike circular geodesics are allowed in the case of equal masses ${\mathcal M}=m$ on the plane $z=b/2$ 
for all values of $\rho $ if  $b>{\tilde b }$, for $\rho\not={\hat \rho }_2 $ if $b={\tilde b }$, and $\rho \in \{[0, {\hat \rho }_2) \cup ({\hat \rho }_3, \infty)\}$ if $b<{\tilde b }$. 
It is interesting to note that when the two masses are placed very close each other ($b\ll 1$) the metric coefficients reduce to
\begin{equation}
\psi(\rho,z)=-\frac{\mathcal{M}+m}{R_C}-\frac{mbz}{R_C^3}\ , \quad \gamma(\rho,z)=-\rho^2 \frac{\mathcal{M}+m}{R_C^4}\left[\frac{\mathcal{M}+m}2+\frac{2zmb}{R_C^2}\right]\ .
\end{equation}
In this approximation timelike circular geodesics are allowed at
\begin{equation}
z=\frac{mb}{\mathcal{M}+m}\ , \quad \rho>2(\mathcal{M}+m)\ .
\end{equation}

For fixed values of $\nu$ geodesics exist only for specific values of $\rho$, such that $a(U)^{\hat \rho}=0$ and $a(U)^{\hat z}=0$.
The condition $a(U)^{\hat z}=0$ requires $\psi_{,z}=0$, and it is satisfied identically in the case of equal masses ${\mathcal M}=m$ on the middle plane $z=b/2$ for all values of $\rho$ (not depending on the chosen value of $\nu$); the condition $a(U)^{\hat \rho}=0$ leads instead to the 
following cubic equation 
\begin{equation}
\label{eqxigeos}
2\nu^2\xi^3-4m(1+\nu^2)\xi^2+mb^2(1+\nu^2)=0\ , \qquad \xi\in\left[b/2, \infty\right)\ ,
\end{equation}
where the new variable $\xi$ defined by $\rho ^2=-b^2/4 + \xi ^2$ is introduced (exactly as we did before). 
In the case $\nu=0$ eq. (\ref{eqxigeos}) reduces to the quadratic equation $-4\xi^2+b^2=0$, whose solution is simply $\xi=\pm b/2$, or $\rho=0$.  
For $\nu\not=0$ the study the solutions of eq. (\ref{eqxigeos}) shows that there exist two positive real roots, which are allowed only for $b\in[0, b_{\rm g})$, with $b_{\rm g}={\tilde b}(1+1/\nu^2)/2$ (and so $b_{\rm g}\in({\tilde b}, \infty)$).
We avoid to give here the corresponding expressions for the roots (which can be easily derived); whereas it is interesting to note that closed loops form around these points when the lines of force of the acceleration field are drawn
for different orbits (see Fig. 7), because these points coincide with the  \lq\lq equilibrium solutions" of the system defining the same lines of force.

\begin{figure} 
\typeout{*** EPS figure 1}
\begin{center}
$\begin{array}{ccc}
\includegraphics[scale=0.25]{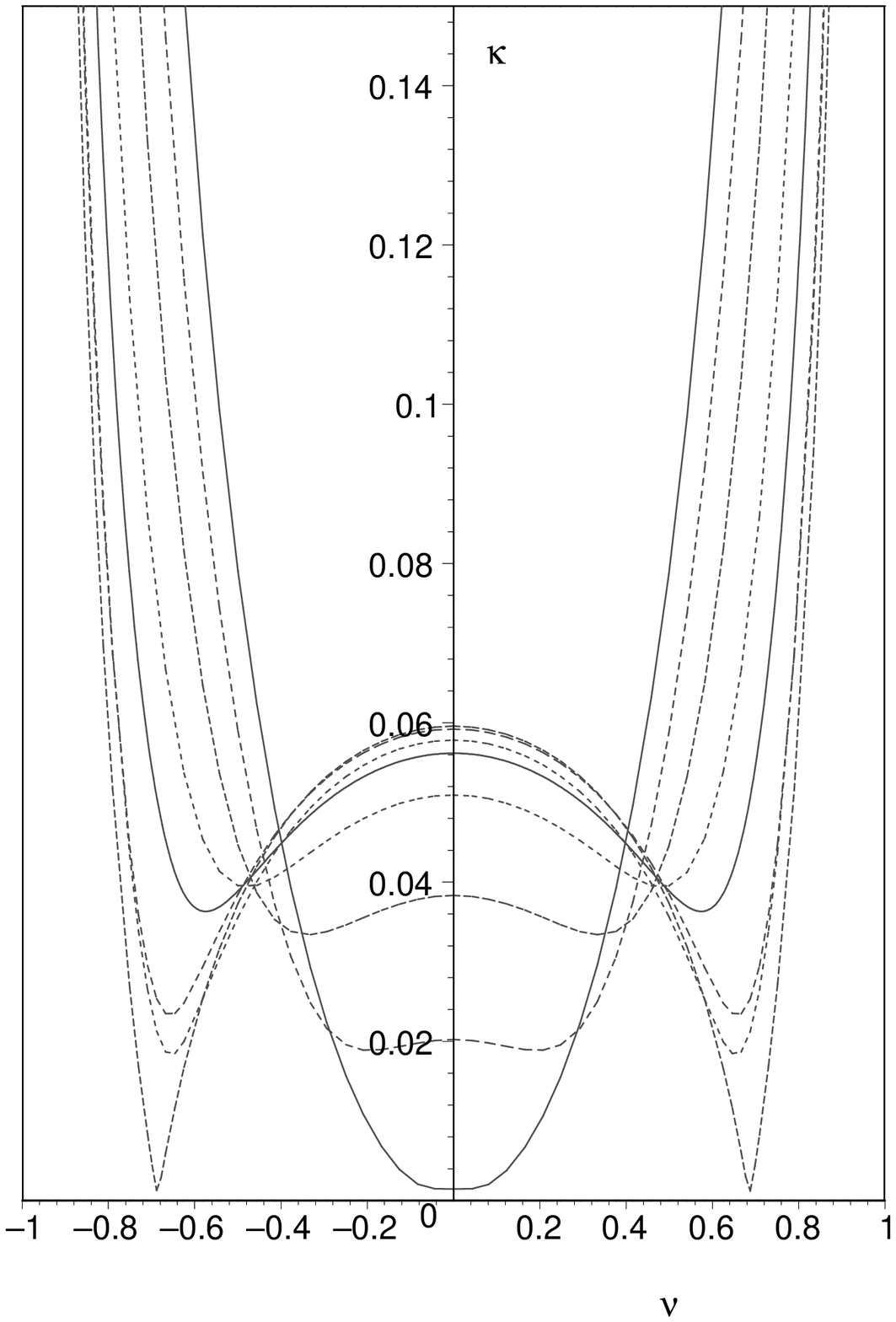}&\qquad
\includegraphics[scale=0.25]{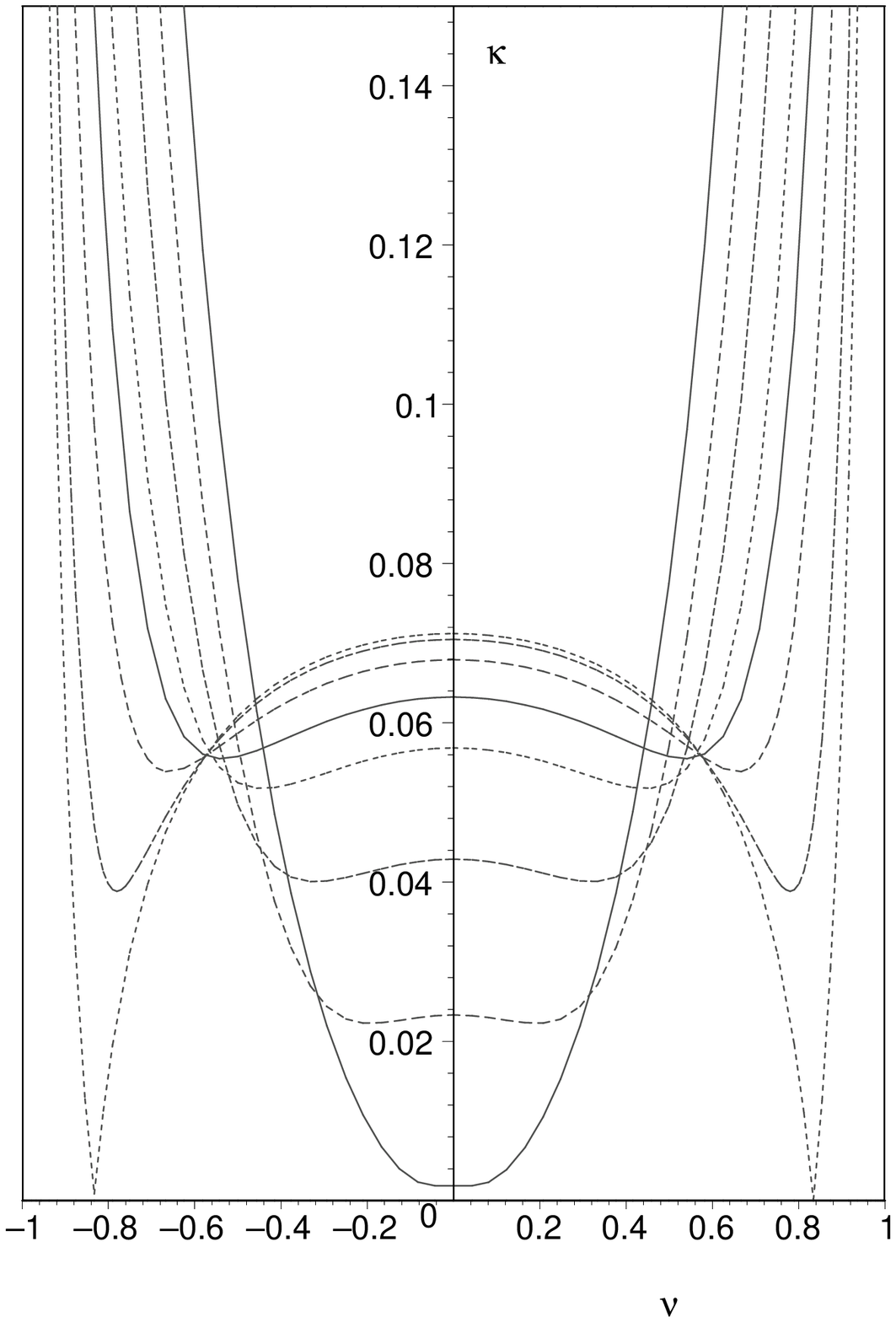}&\qquad
\includegraphics[scale=0.25]{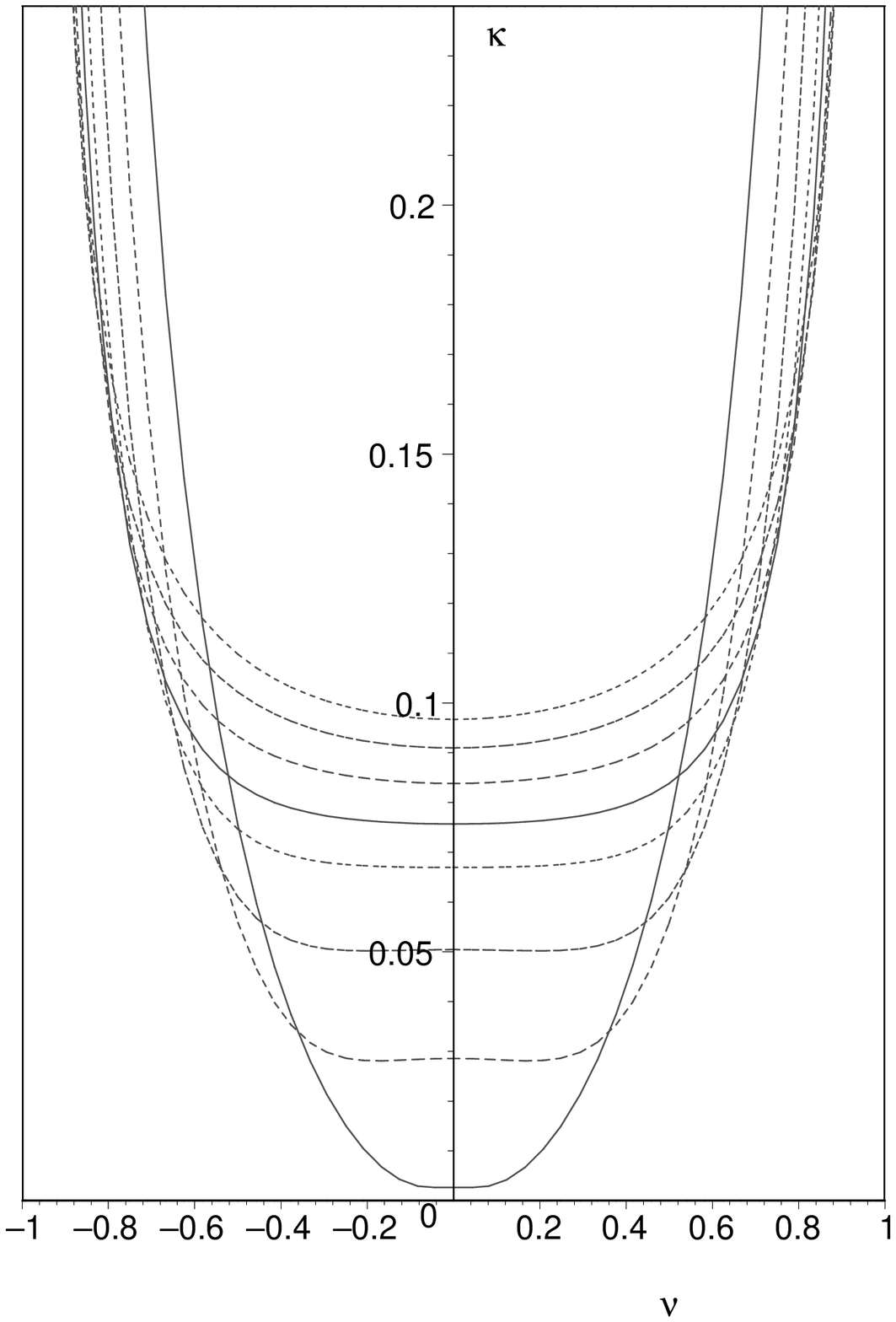}\\
\mbox{(a)} &\qquad \mbox{(b)}& \qquad\mbox{(c)}
\end{array}
$\\
\end{center}
\caption{The magnitude of the acceleration $\kappa$ as a function of the linear velocity $\nu$, for ${\mathcal M}=1$, $b=3$ and different values of $m=1/2, 1, 2$, for Fig. (a), (b) and (c) respectively. The curves are obtained by fixing the coordinate $\rho=4$, and $z$ assuming the values $[-10b, -2b, -b, -b/2,-b/4, 0, b/4, b/2]$ (from bottom to top). Note that only in the case of equal masses of case (b), the curve corresponding to $z=b/2$ really intersects the $\kappa=0$ axis.}  
\label{fig:1}
\end{figure}


\begin{figure}
\typeout{*** EPS figure 2}
\begin{center}
$\begin{array}{ccc}
\includegraphics[scale=0.25]{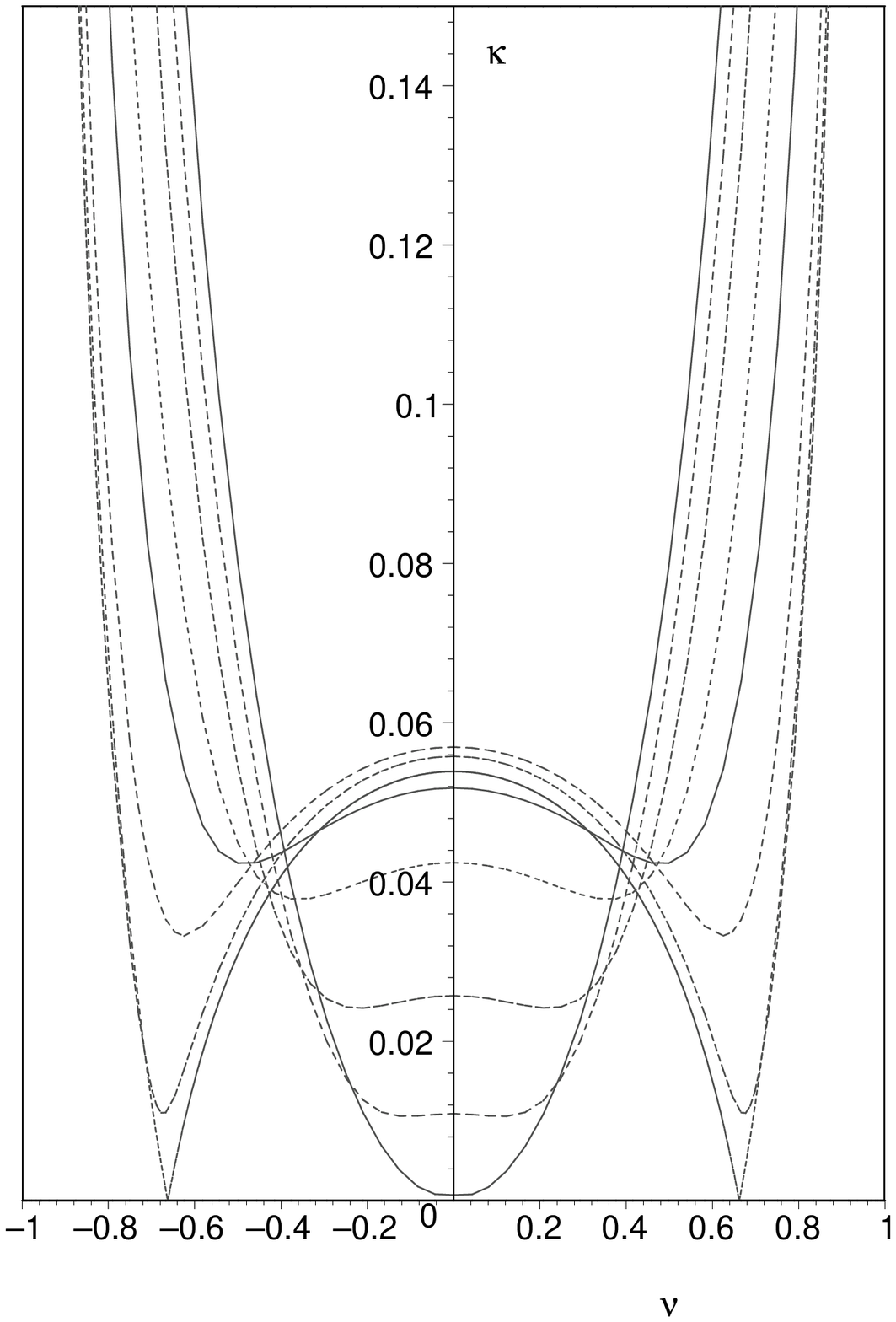}&\qquad
\includegraphics[scale=0.25]{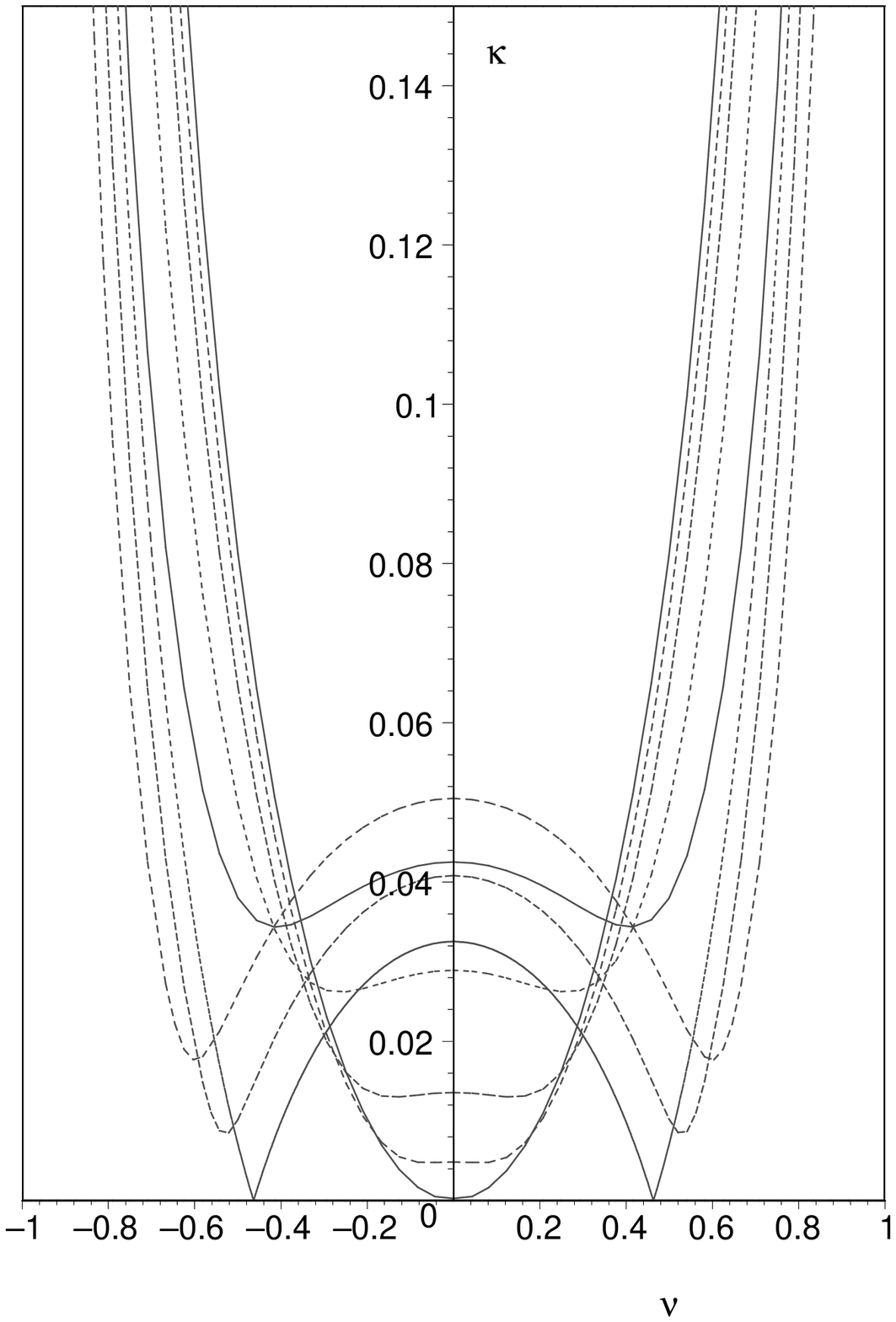}&\qquad
\includegraphics[scale=0.25]{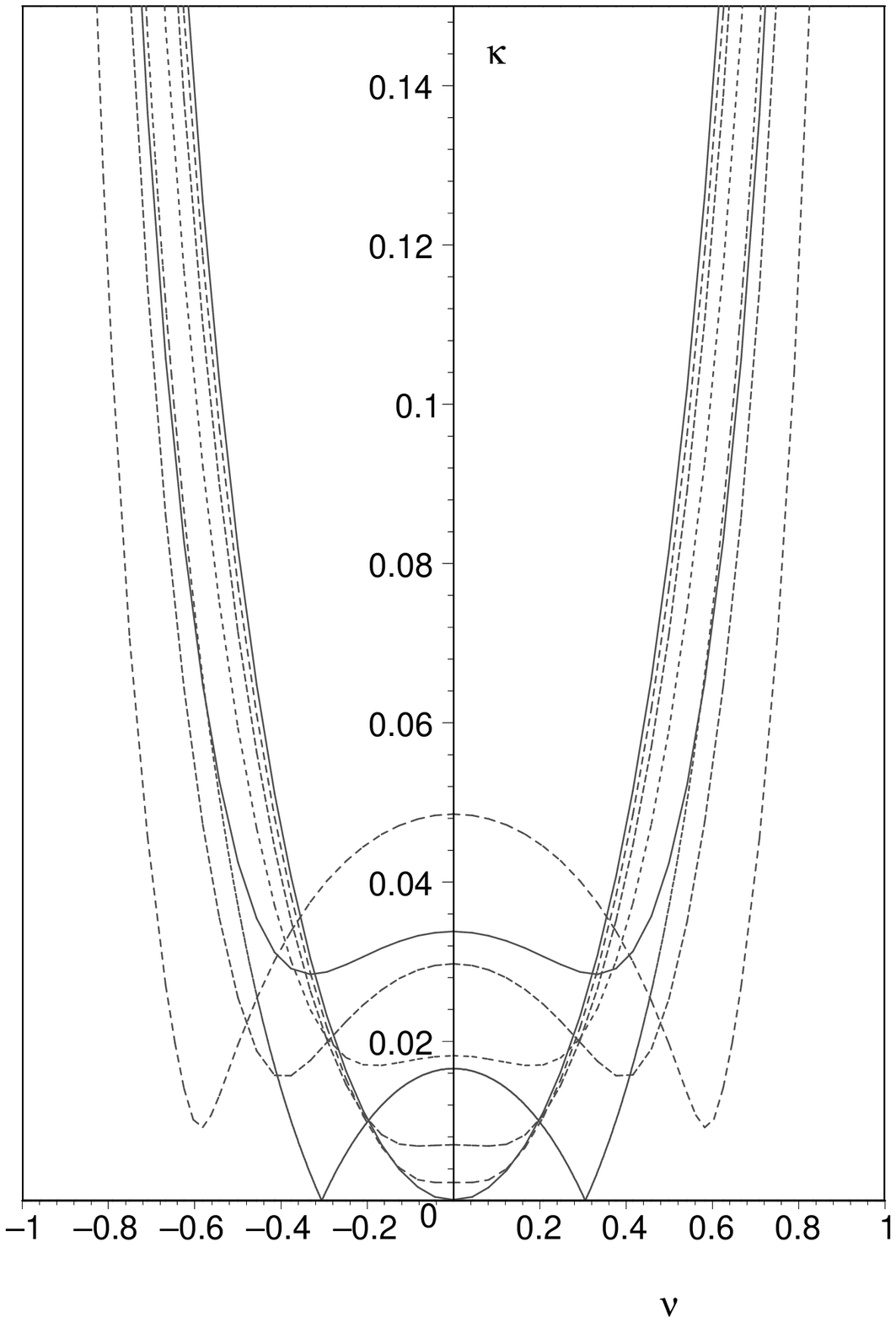}\\
\mbox{(a)} &\qquad \mbox{(b)}&\qquad  \mbox{(c)}
\end{array}$\\
\end{center}
\caption{The behavior of $\kappa$ as a function of the linear velocity $\nu \in (-1, 1)$ is shown for different values of the parameter $b=5, 8, 12$ respectively, in the case $m={\mathcal M}=1$. 
The curves refer to the same points $(\rho ,z)$ of Fig. \ref{fig:1}.}  
\label{fig:2}
\end{figure}


\begin{figure}
\typeout{*** EPS figure 3}
\begin{center}
$\begin{array}{ccc}
\includegraphics[scale=0.25]{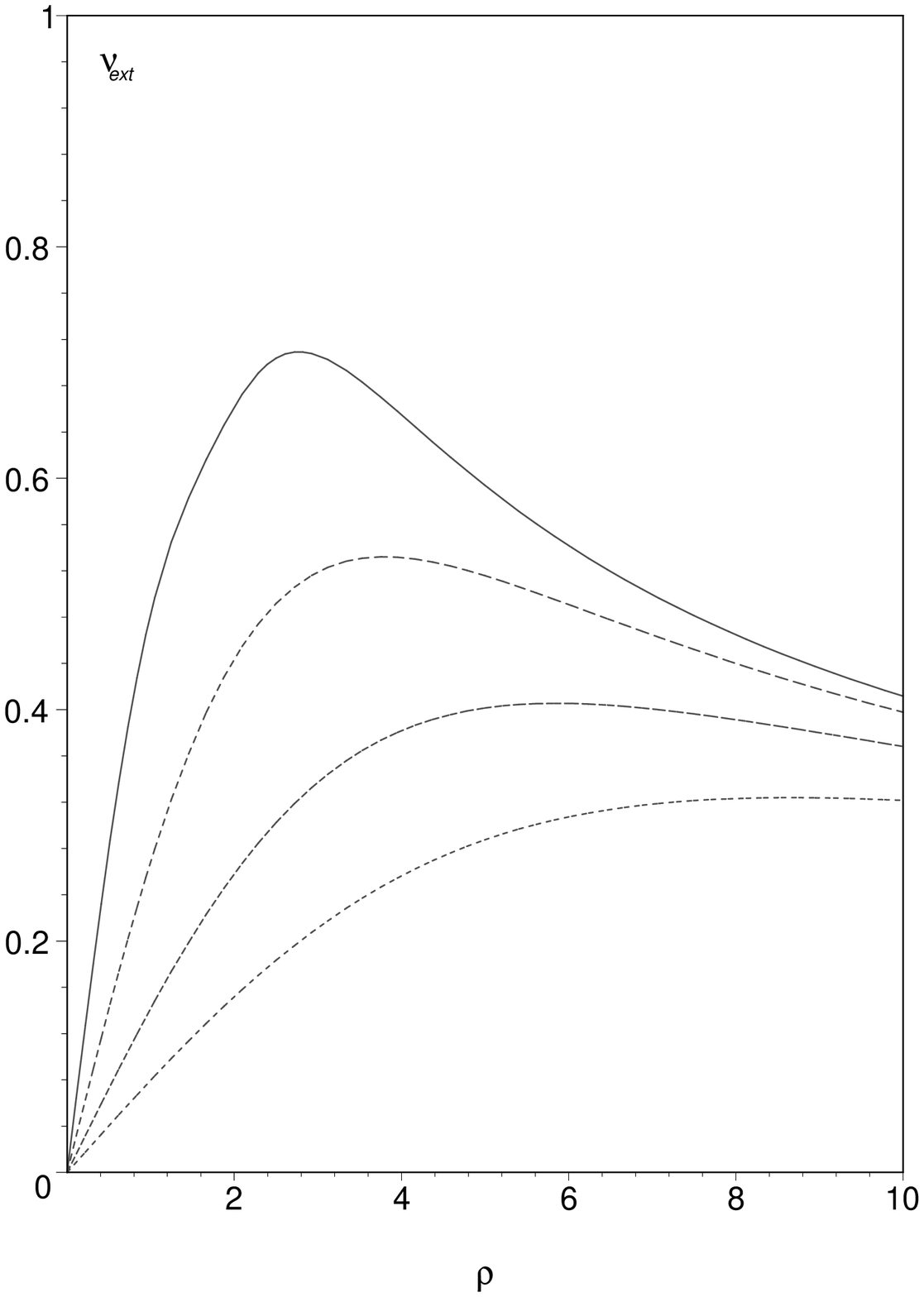}&\qquad
\includegraphics[scale=0.25]{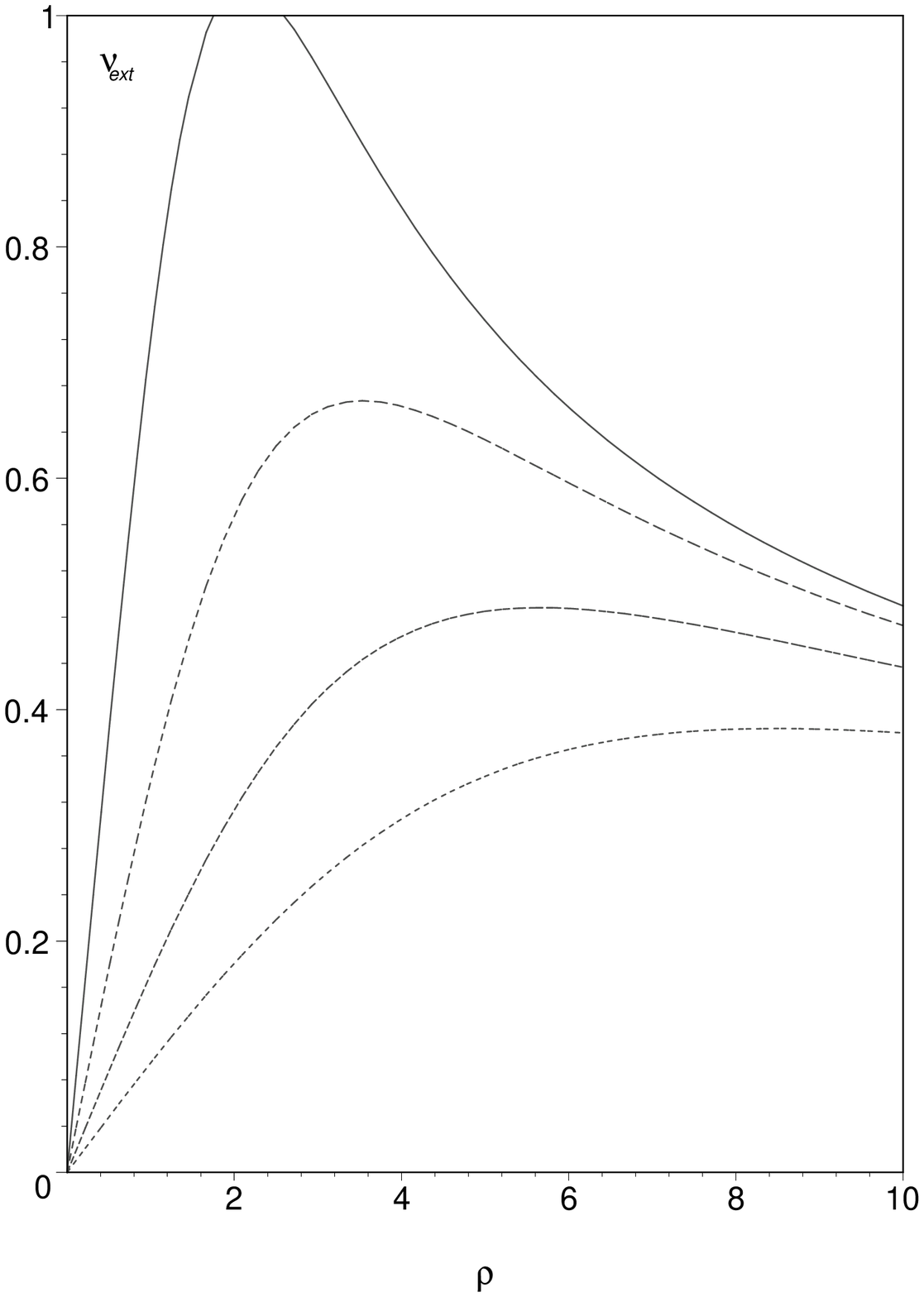}&\qquad
\includegraphics[scale=0.25]{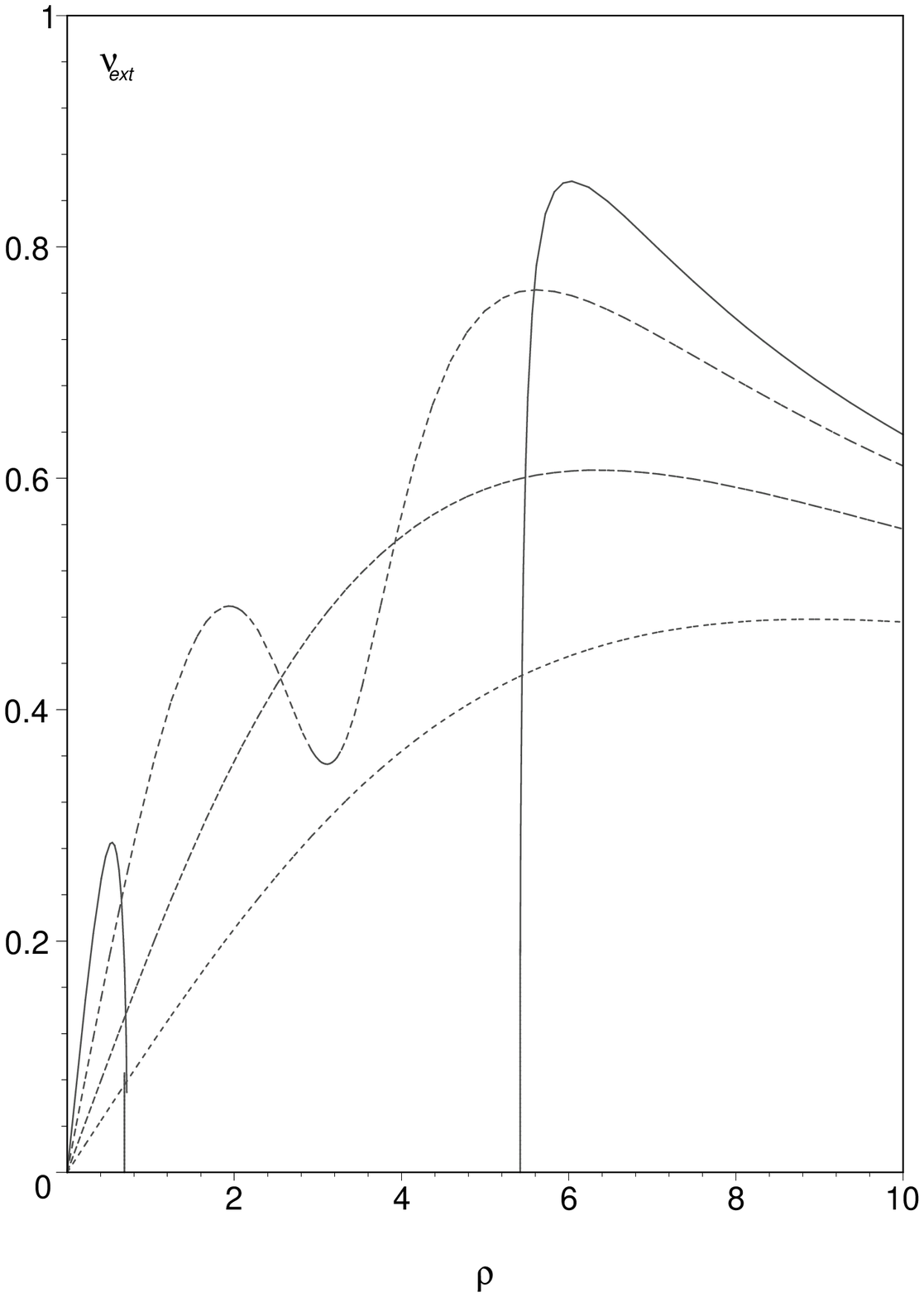}\\
\mbox{(a)} &\qquad \mbox{(b)}&\qquad  \mbox{(c)}
\end{array}$\\
\end{center}
\caption{The relative velocity $\nu_{(\kappa, \rm ext)}{}_{(+)}$ corresponding to the (co-rotating) extremely accelerated orbits is plotted as a function of the coordinate $\rho$, for fixed values of $z=b/2$, $b=3, 5, 8, 12$ (from top to bottom) and ${\mathcal M}=1$. 
Figs. (a), (b) and (c) correspond to ${\mathcal M}=1$ and different values of  $m=1/2, 1, 2$ respectively.
Note that for certain selection of parameters (the case $m>M$  of Fig. (c))  $\nu_{(\kappa, \rm ext)}{}_{(+)}$  exhibits very different behaviors: large values of $b$ correspond to the curves similar to those of case (a) and (b); when $b$ decreases ($b=5$ is the value chosen in the graphics) one starts seeing oscillations which finally degenerate in a forbidden region for $\nu_{(\kappa, \rm ext)}{}_{(+)}$ (already present when $b=3$). This is simply explained by looking at extremal points of the graphics in Figs. 1 and 2.  
}  
\label{fig:3}
\end{figure}


\begin{figure} 
\typeout{*** EPS figure 4}
\begin{center}
$\begin{array}{ccc}
\includegraphics[scale=0.25]{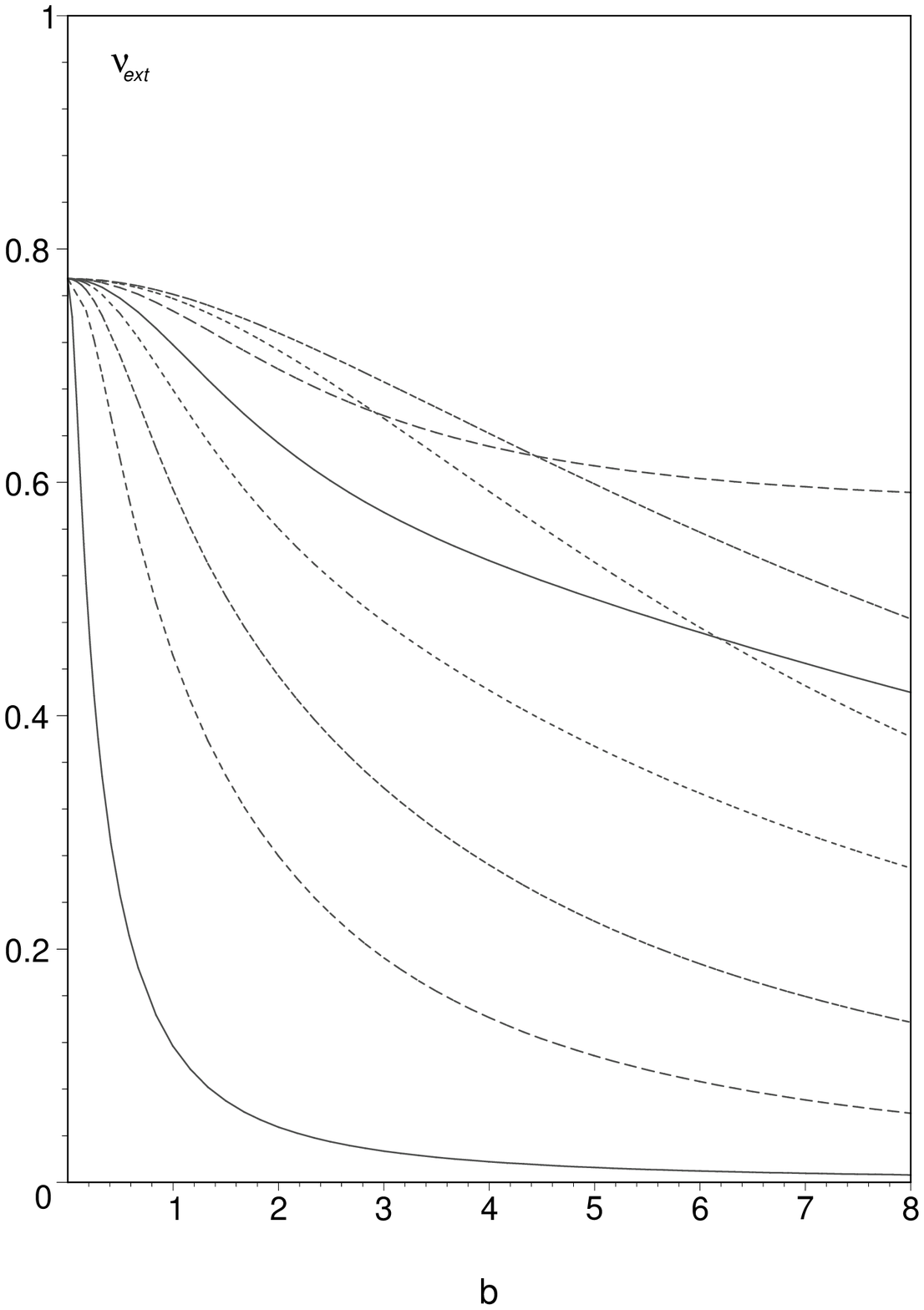}&\qquad
\includegraphics[scale=0.25]{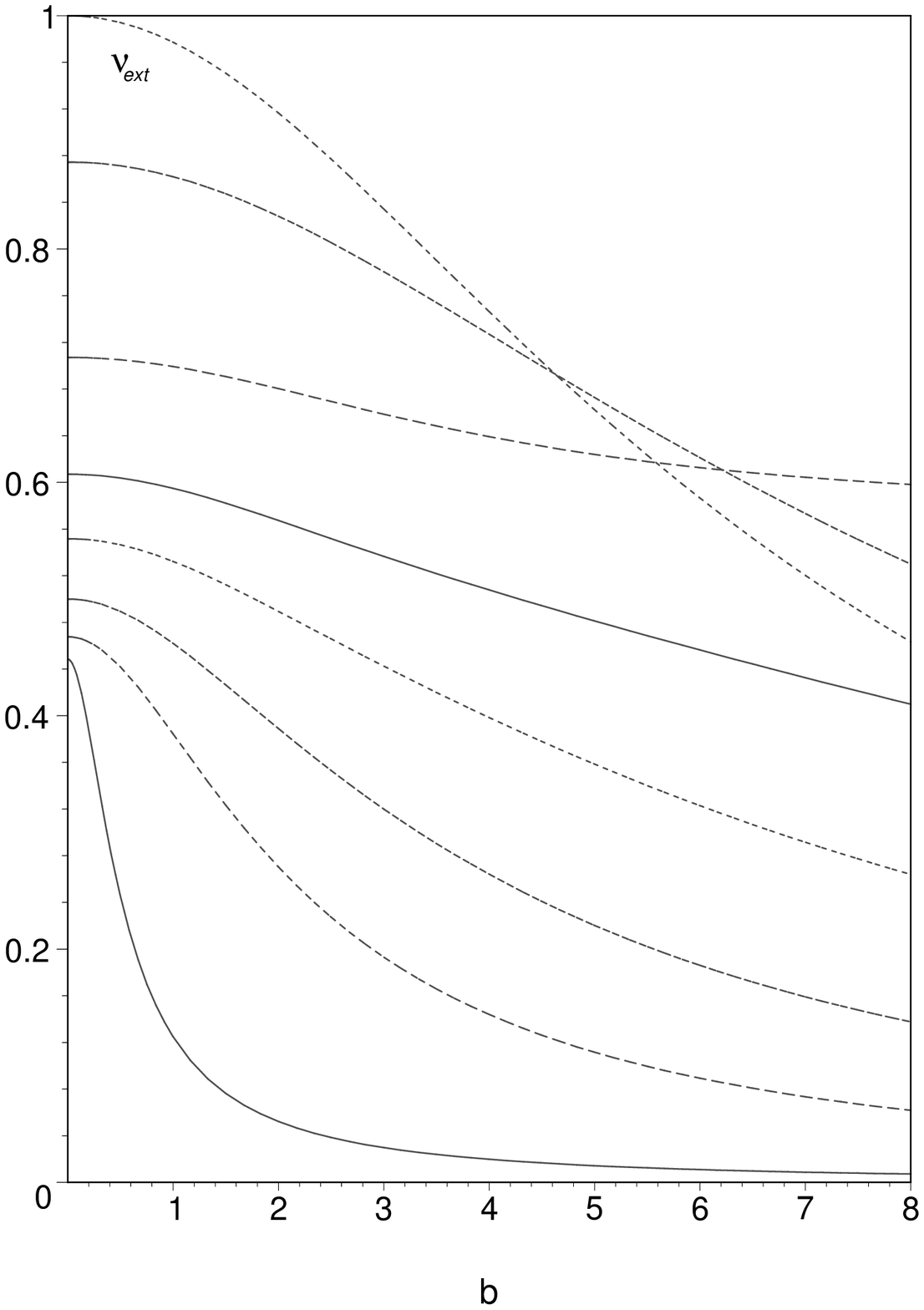}&\qquad
\includegraphics[scale=0.25]{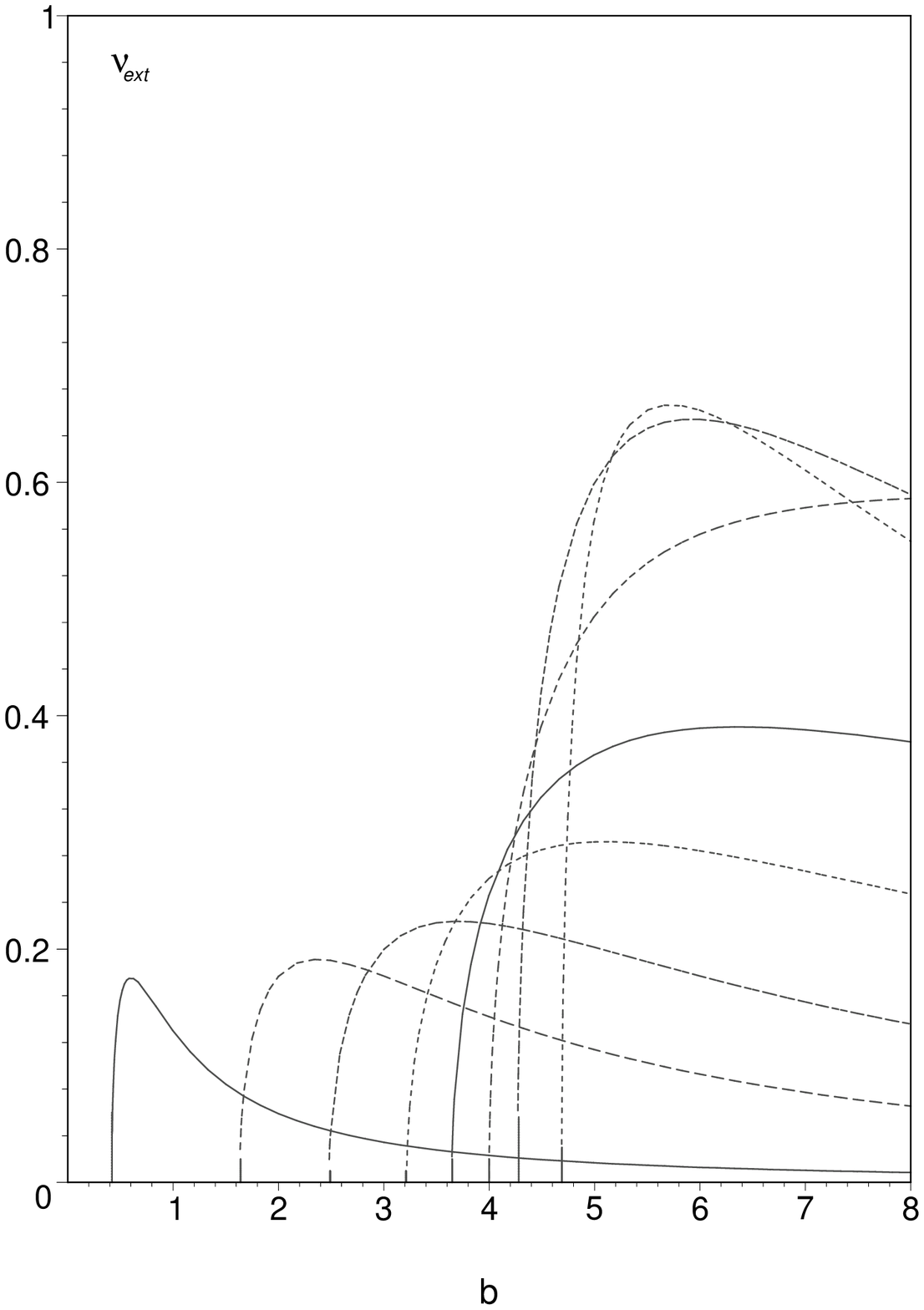}\\
\mbox{(a)} &\qquad \mbox{(b)}&\qquad  \mbox{(c)}
\end{array}$\\
\end{center}
\caption{Behavior of $\nu_{(\kappa, \rm ext)}{}_{(+)}$ as a function of the separation parameter $b$ for ${\mathcal M}=1$ and for $m=1/2, 1, 2$ respectively.
The curves (from bottom to top) refer to the same points $(\rho ,z)$ of Fig. \ref{fig:1}.}  
\label{fig:4}
\end{figure}


\begin{figure}
\typeout{*** EPS figure 5}
\begin{center}
$\begin{array}{ccc}
\includegraphics[scale=0.25]{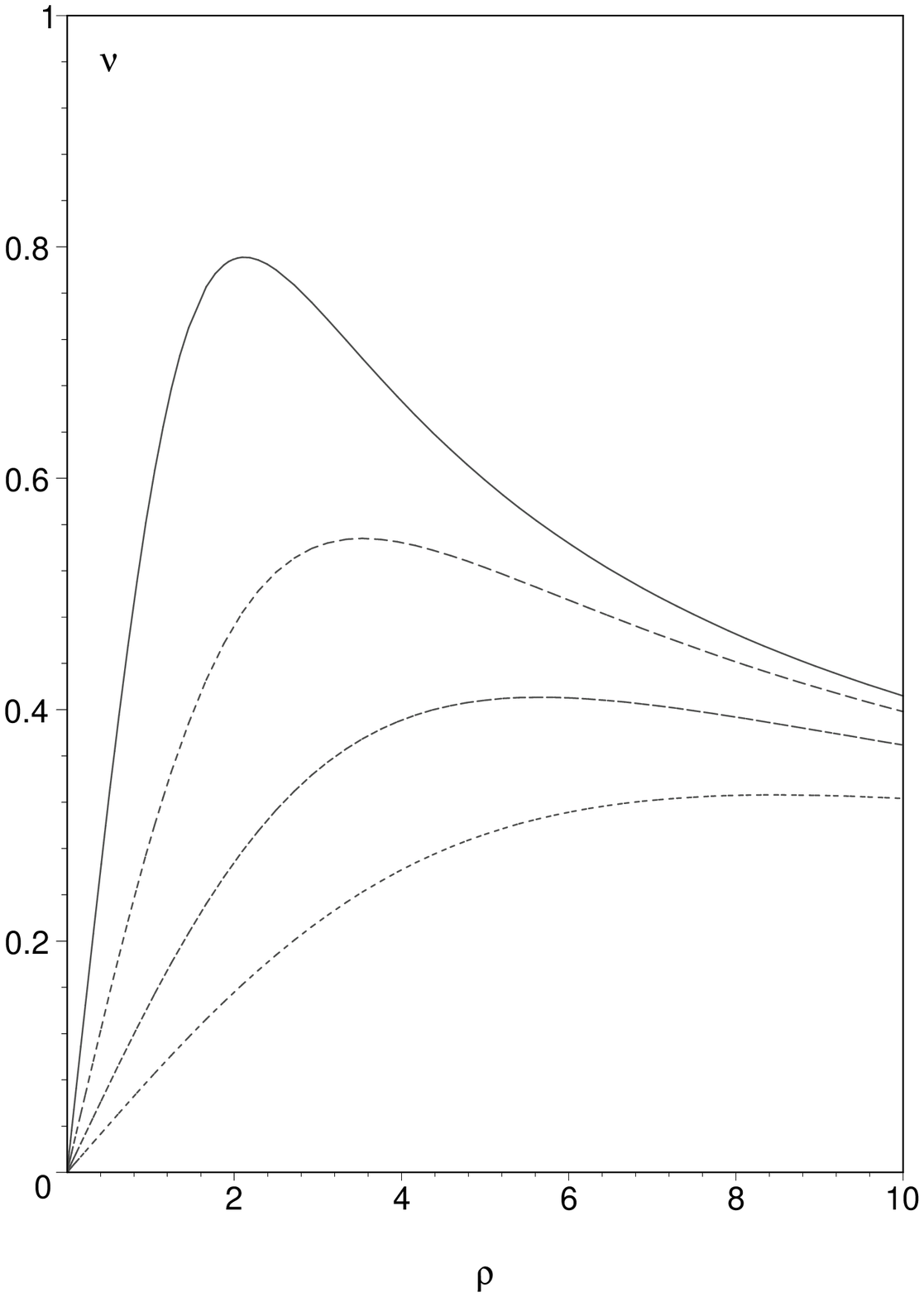}&\qquad
\includegraphics[scale=0.25]{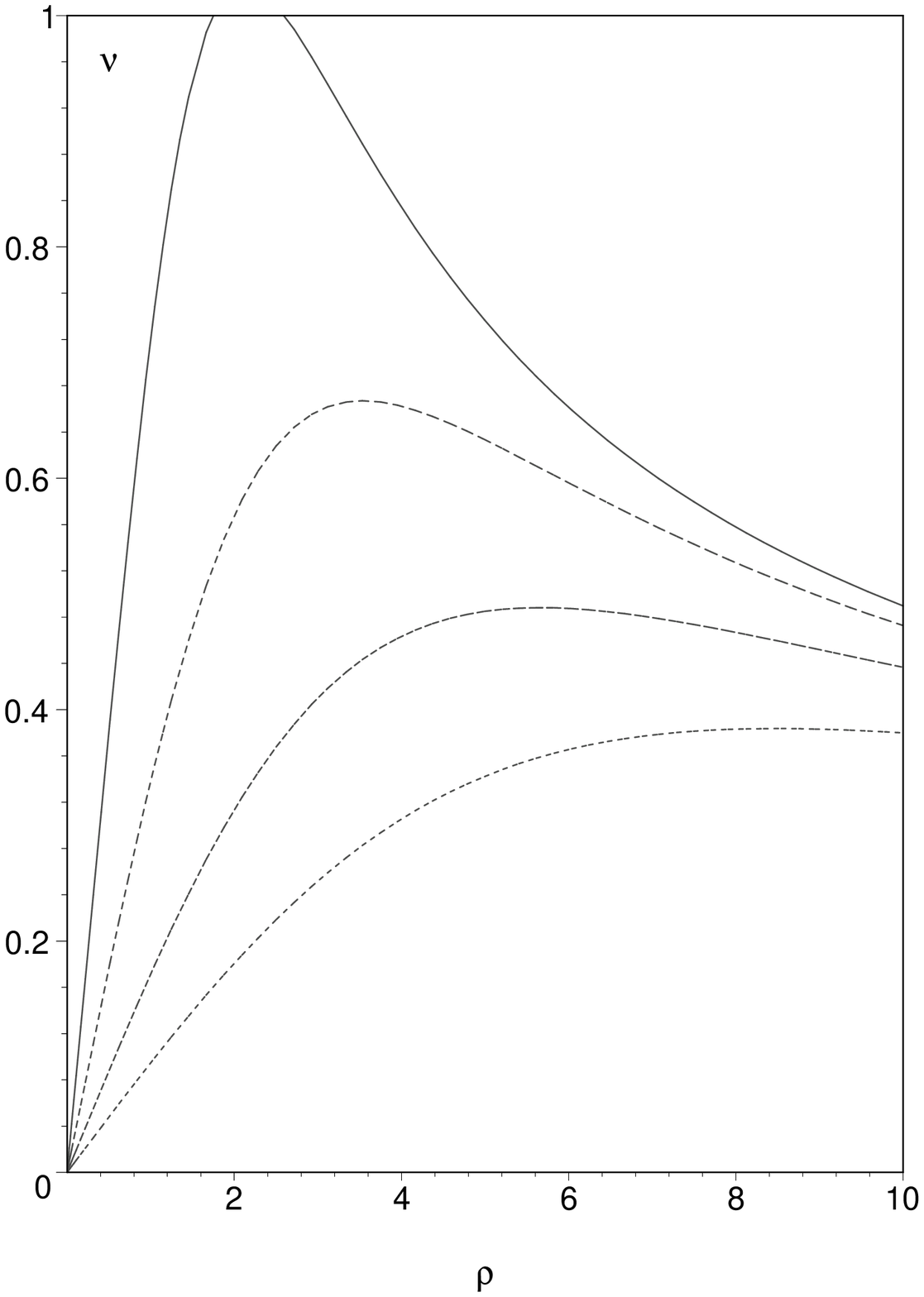}&\qquad
\includegraphics[scale=0.25]{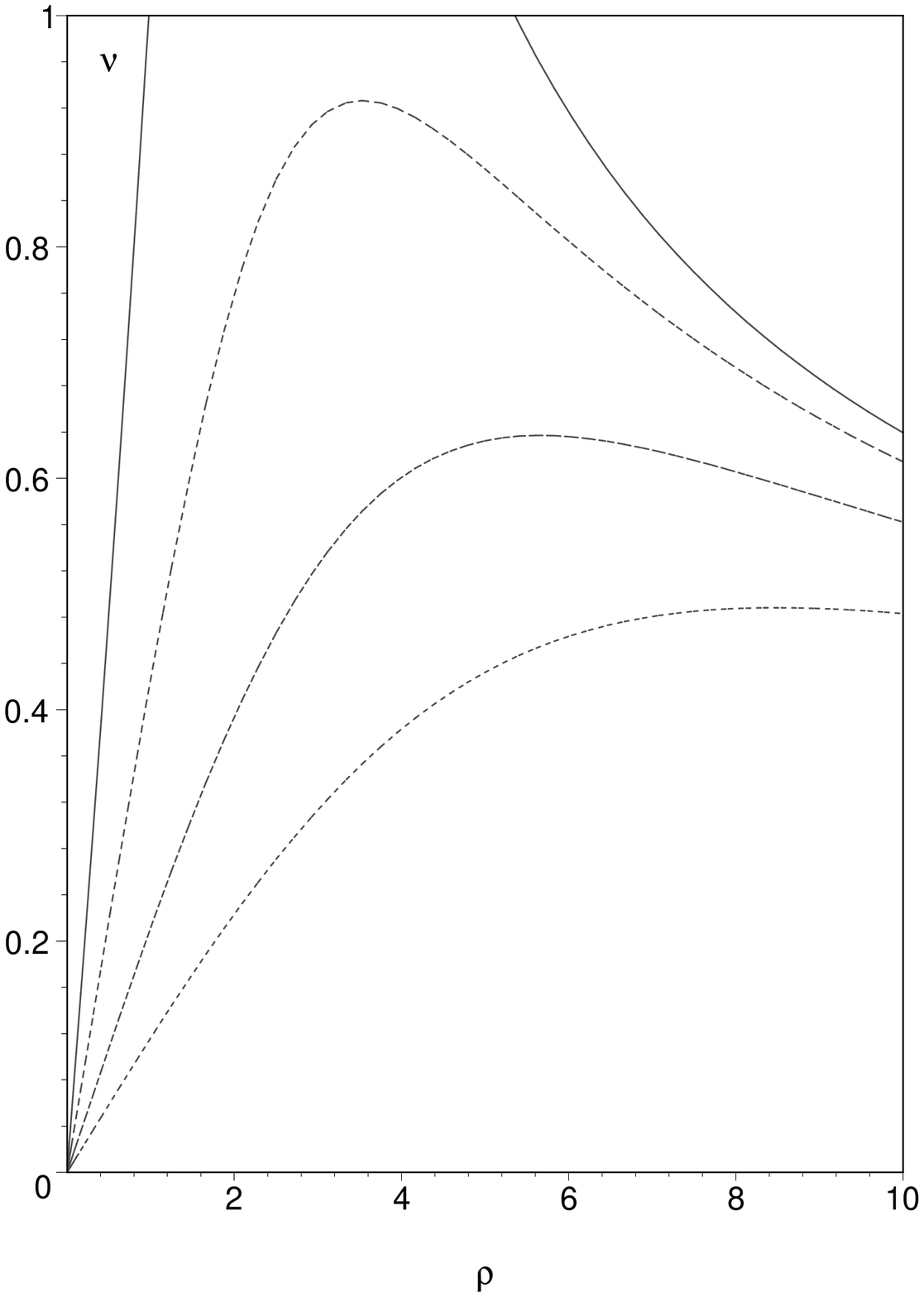}\\
\mbox{(a)} &\qquad \mbox{(b)}&\qquad  \mbox{(c)}
\end{array}$\\
\end{center}
\caption{The geodesic relative velocity $\nu_+$ is plotted as a function of the coordinate $\rho$, for fixed values of $z=b/2$, $b=3, 5, 8, 12$ (from top to bottom) and for ${\mathcal M}=1$. 
Figs. (a), (b) and (c) correspond to different values of  $m=1/2, 1, 2$ respectively.}  
\label{fig:5}
\end{figure}


\begin{figure} 
\typeout{*** EPS figure 6}
\begin{center}
$\begin{array}{ccc}
\includegraphics[scale=0.25]{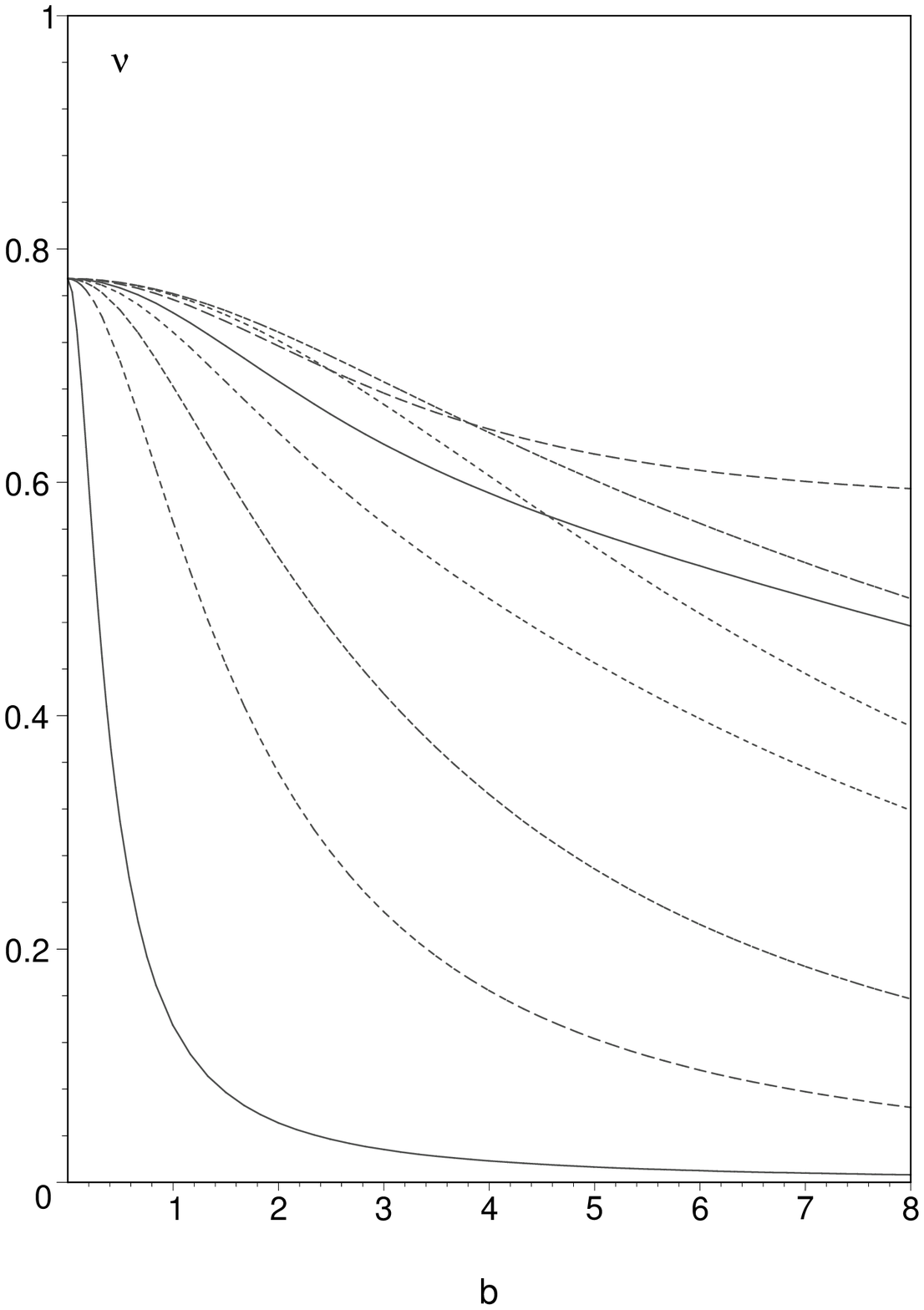}&\qquad
\includegraphics[scale=0.25]{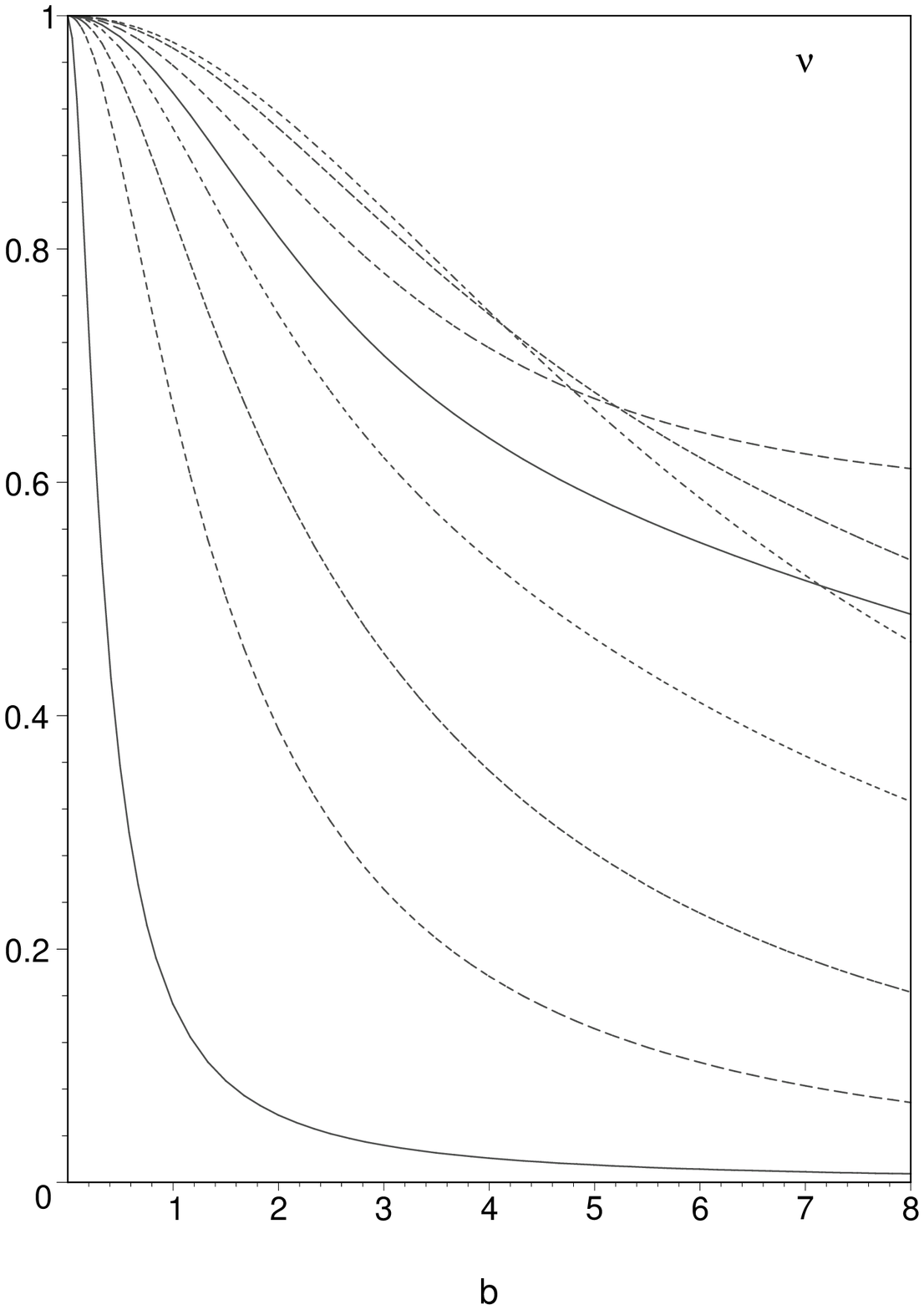}&\qquad
\includegraphics[scale=0.25]{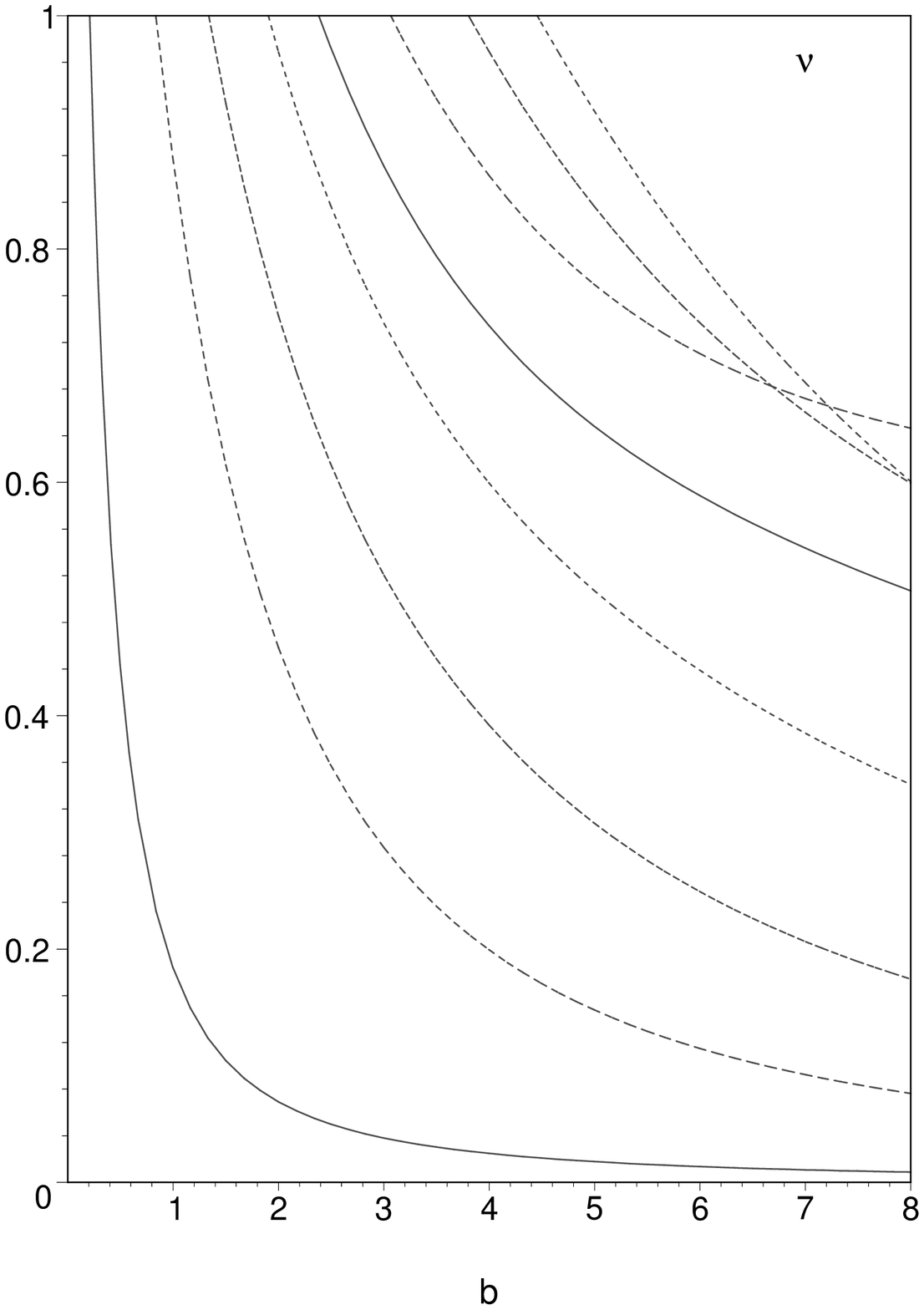}\\
\mbox{(a)} &\qquad \mbox{(b)}&\qquad  \mbox{(c)}
\end{array}$\\
\end{center}
\caption{Behavior of $\nu_+$ as a function of the separation parameter $b$ for ${\mathcal M}=1$ and different values for $m=1/2, 1, 2$ respectively.
The curves (from bottom to top) refer to the same points $(\rho ,z)$ of Fig. \ref{fig:1}. Orbits with $\nu=\nu_+$ are accelerated along $z$.}  
\label{fig:6}
\end{figure}


\begin{figure}
\typeout{*** EPS figure 7}
\begin{center}
$\begin{array}{c@{\hspace{1in}}c}
\includegraphics[scale=0.3]{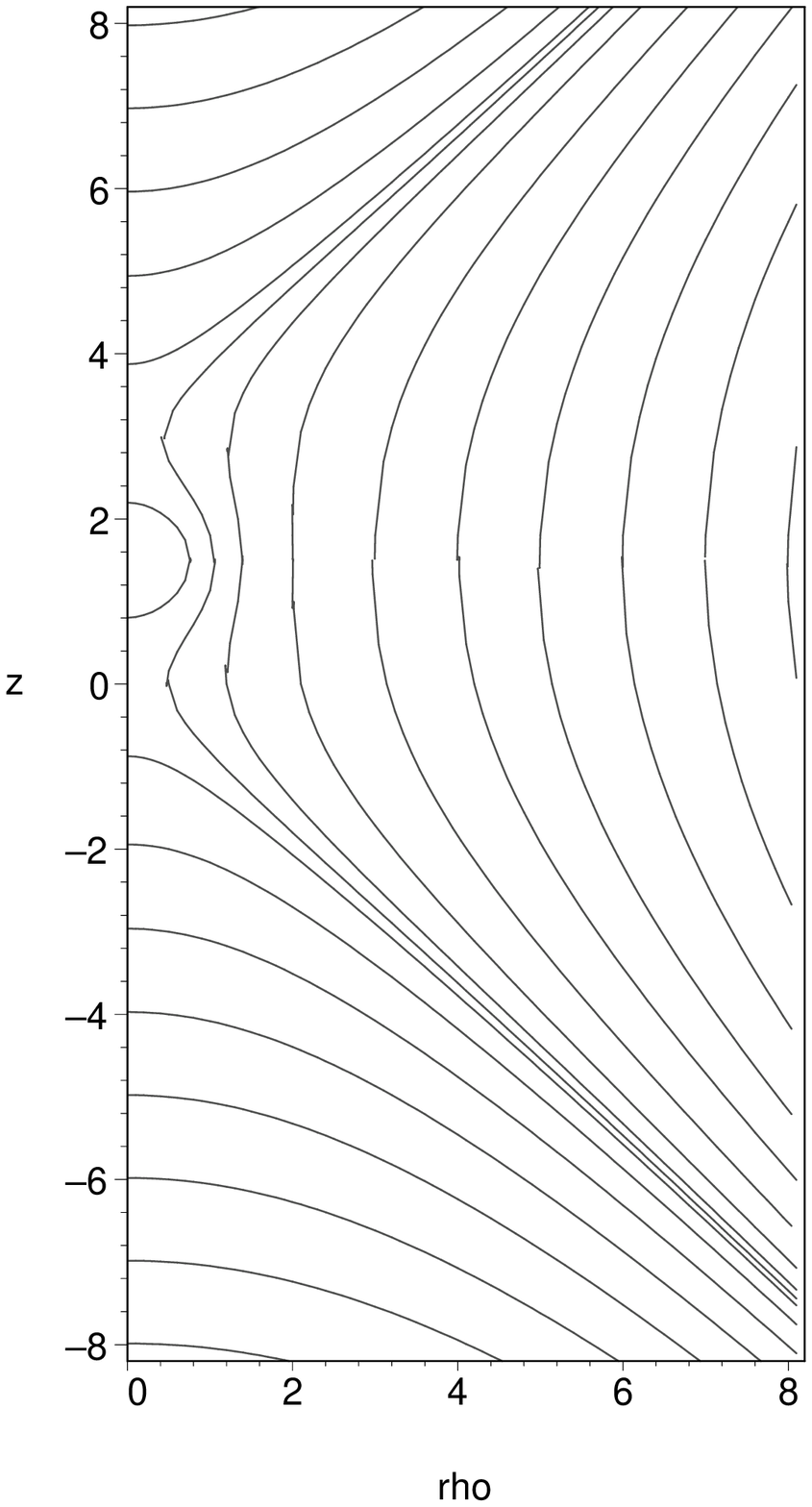}&
\includegraphics[scale=0.3]{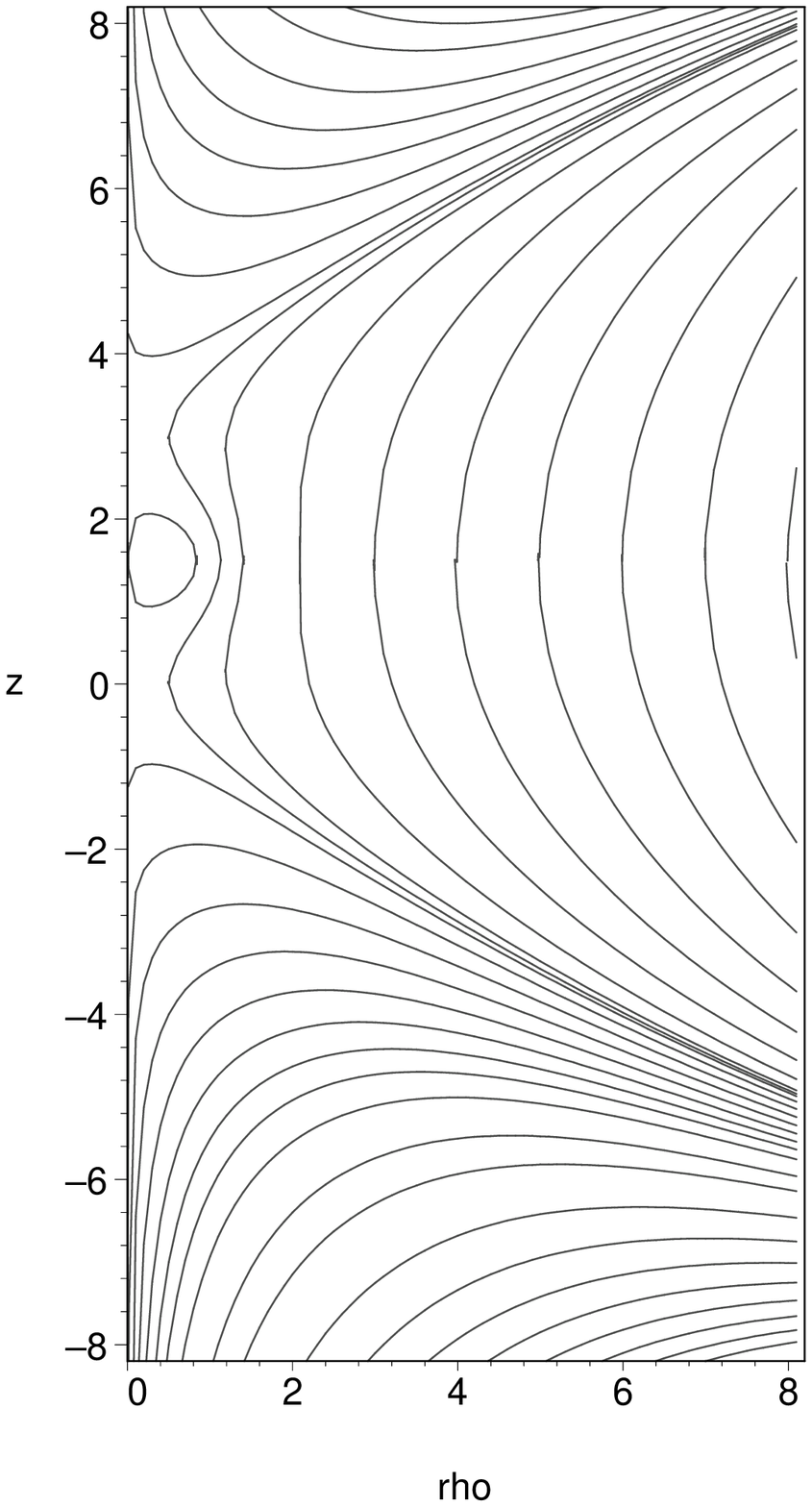}\\[0.4cm]
\mbox{(a)} & \mbox{(b)}\\[0.6cm]
\includegraphics[scale=0.3]{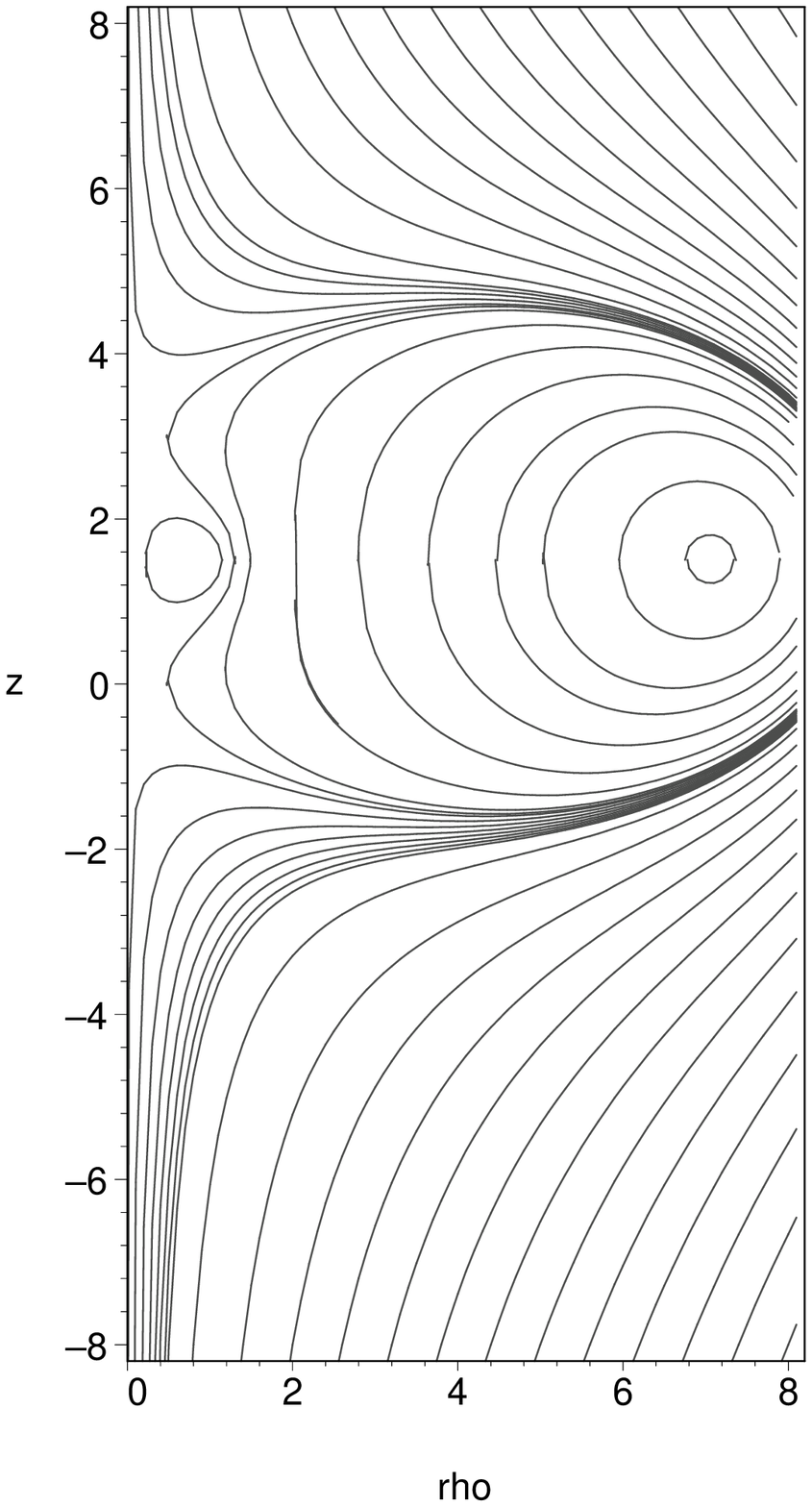}&
\includegraphics[scale=0.3]{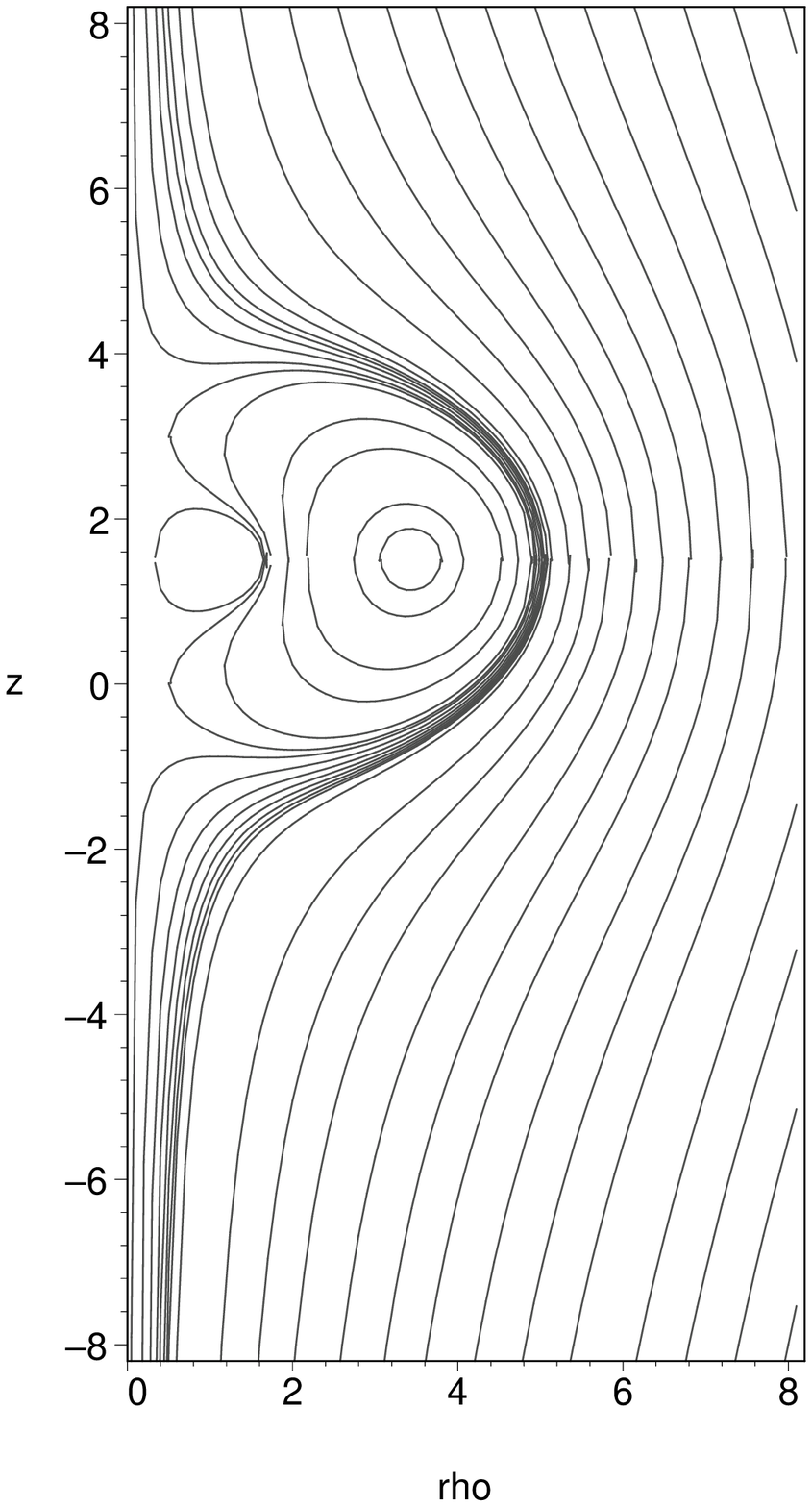}\\[0.4cm]
\mbox{(c)} & \mbox{(d)}\\
\end{array}$\\
\end{center}
\caption{
The lines of force of the acceleration field for different orbits corresponding to $\nu=0,0.3,0.6,0.9$ (Fig. (a), (b), (c), (d), respectively)
with  $b=3$ and ${\mathcal M}=1=m$ are plotted. 
One can easily show that the request ${\mathcal M}=1=m$ implies that there exist certain \lq\lq equilibrium points'' around which closed loops form in the case $\nu\not=0$. As a consequence of the previous choice of the parameters, it can be shown by solving eq. (\ref{eqxigeos}) that they are located at $z=b/2$ and  
$\rho_1\approx 0.3923$ and $\rho_2\approx 24.0819$ in case (b) (even if the second point is out of the range of values which are plotted), $\rho_1\approx 0.8096$ and $\rho_2\approx 7.0731$ in case (c), $\rho_1\approx 1.3689$ and $\rho_2\approx 3.4441$ in case (d).
It is worth noticing that the behavior of the lines of force when $\nu\gg 1$ 
does not show significative variations in comparison with Fig. 7 (d), so that this figure qualitatively represents also the lines of force of the Lie relative curvature $k_{\rm (lie)}$ which are formally obtained from those of $a(U)$ taking the limit $\nu\to \infty$.}  
\label{fig:7}
\end{figure}

Interesting cases correspond to different masses ($m\not={\mathcal M}$) for $b,\rho$ fixed, and $z$ varying values, in comparison with the (symmetric) equal masses configuration.  
Two oppositely rotating circular geodesics are allowed at $z=b/2$ under the assumption $m={\mathcal M}$ (see Fig. 1 (b)), but this is no more true when the masses are different (see Fig. 1 (a), where $m=1/2$ and ${\mathcal M}=1$ and (c), where $m=2$ and ${\mathcal M}=1$).
Note that in Fig. 1 (a) and (c), because of the different values for the masses, $\kappa$ never vanishes in spite of the apparent similarity of the graphics  in Fig. 1 (a) and (b).

The behavior of $\kappa (\nu) $ on a $z=const$ plane presents two interesting situations: very far from the two particles and approaching the location of a single particle. In the first case, the curve presents a single (local) minimum at $\nu=0$ which, in the second one, becomes a (local) maximum, while two other minima appear in symmetric positions for values of $\nu$ approaching the speed of light. 
This picture has now a clear relativistic explanation \cite{idcf2}. In fact, far from the system of two particles, one is left with the Newtonian situation only: increasing the speed, the four acceleration increases, and hence the force needed to maintain the orbit increases too. 
Approaching one of the particles, relativity makes things no more intuitive: 
increasing the speed from zero to the value corresponding to the position of the (newly generated local) minima, implies that the four acceleration decreases and also the force necessary to maintain the orbit. 
In other words, we are in presence of the so called \lq\lq centrifugal force reverse'' \cite{idcf1,idcf2,abram}, which is now seen in a different context with respect to that of black holes where it has been first introduced. 
As soon as the value of $z$ reaches $b/2$, in the case of equal masses, the two local minima correspond to $\kappa=0$, i.e. they identify the geodesics. 
The discussion for all possible values of $z$ is obtained by symmetry considerations so that the only range to be explored in the case $m={\mathcal M}$ is $z\in (-\infty, b/2]$. 
The study of $\kappa$ as a function of $b$ is investigated in Fig. 2 (a), (b), (c) in the case of equal masses.
In Fig. 3 (a), (b) and (c) $\nu_{( \kappa, \rm ext)}{}_{(+)}$ is plotted with a special choice of parameters as a function of $\rho$. 
Note that for certain selection of parameters (the case $m>M$  of Fig. 3(c))  $\nu_{(\kappa, \rm ext)}{}_{(+)}$  exhibits very different behaviors: large values of $b$ correspond to the curves similar to those of case (a) and (b); when $b$ decreases ($b=5$ in the graphics) one starts seeing oscillations which then degenerate in a forbidden region for $\nu_{(\kappa, \rm ext)}{}_{(+)}$ (already present when $b=3$). In Fig. 4 (a), (b) and (c) $\nu_{( \kappa, \rm ext)}{}_{(+)}$ is plotted with a special choice of parameters as a function of $b$. 
In Fig. 5 and 6 one has the same of Fig. 3 and 4 repeated for the geodesics $\nu_{\pm}$  relative velocities (actually in the graphics is indicated only $\nu_+$, being $\nu_-=-\nu_+$).

In Fig. 7 the lines of force of the acceleration field for different orbits, corresponding to different values of $\nu$, are drawn in the case of equal masses $m={\mathcal M}$ and for a fixed value of the separation parameter $b<b_{\rm g}$ (in such a way timelike circular geodesics are allowed to exist). When $\nu\not = 0$ closed loops form around certain \lq\lq equilibrium points'' corresponding to timelike circular geodesics (see Figs. \ref{fig:7} (b), (c), (d)). These are instead absent in the case $\nu= 0$, represented in Fig. \ref{fig:7} (a), where the lines of force, from their defining condition, are
\begin{equation}
\frac{\rmd \rho}{{\rm d} z}=\frac{\psi_{,\rho}}{\psi_{,z}}
\end{equation}
and they coincide with the lines of force of the vector field orthogonal to the surface $\psi=const$.
Note that in the lines of force
the effects of a conical singularity along the axis disappear (the conical singularity infact is due to a nonzero limiting value for $\gamma$ on the $z-$axis but in the defintion (\ref{linesofforce}) $\gamma$ cancels out).

\subsection{Superposition of three Chazy-Curzon particles}

The solution corresponding to the superposition of three Chazy-Curzon particles with masses ${\mathcal M}$, $m$, $\mu$ and positions $z=0$, $z=b$ and $z=c$ (with $c>b$) on the $z-$axis respectively is given by metric (\ref{weylmetric}) with functions
\begin{eqnarray}
\label{psigammaCCC}
\psi=\psi_{\rm C}+\psi_{\rm C_b}+\psi_{\rm C_c}\ , \qquad \gamma=\gamma_{\rm C}+\gamma_{\rm C_b}+\gamma_{\rm C_c}+\gamma_{\rm CCC}\ ,
\end{eqnarray}
where $\psi_{\rm C}$, $\psi_{\rm C_b}$ and $\gamma_{\rm C}$, $\gamma_{\rm C_b}$ are given by (\ref{ccsolw}), and
\begin{eqnarray}
\psi_{\rm C_c}=-\frac{\mu}{R_{\rm C_c}}\ , \qquad \gamma_{\rm C_c}=-\frac12\frac{\mu^2\rho^2}{R_{\rm C_c}^4}\ , \qquad R_{\rm C_c}=\sqrt{\rho^2+(z-c)^2}\ 
\end{eqnarray}
while $\gamma_{\rm CCC}$ can be obtained by solving Einstein's equations (\ref{einsteqs}):
\begin{eqnarray}
\gamma_{\rm CCC}&=&2\frac{m{\mathcal M}}{b^2}\frac{\rho^2+z(z-b)}{R_{\rm C}R_{\rm C_b}}+2\frac{\mu{\mathcal M}}{c^2}\frac{\rho^2+z(z-c)}{R_{\rm C}R_{\rm C_c}}\nonumber\\
&&+2\frac{\mu m}{(b-c)^2}\frac{\rho^2+(z-b)(z-c)}{R_{\rm C_b}R_{\rm C_c}}+C\ .
\end{eqnarray}
The value of arbitrary constant $C$ can be determined by imposing the regularity condition (\ref{regcond}), from which we have
\begin{eqnarray}
0=\lim_{\rho\rightarrow0}\gamma_{\rm CCC}&=&2\frac{m{\mathcal M}}{b^2}\frac{z(z-b)}{|z||z-b|}+2\frac{\mu{\mathcal M}}{c^2}\frac{z(z-c)}{|z||z-c|}\nonumber\\
&&+2\frac{\mu m}{(b-c)^2}\frac{(z-b)(z-c)}{|z-b||z-c|}+C\ .
\end{eqnarray}
The relevant cases to analyze are the following:
\begin{itemize}
\item{case 1: $z>c$ (and, analogously, $z<0$):
\begin{equation}
0=\gamma_{\rm CC}(0,z)=2\frac{m{\mathcal M}}{b^2}+2\frac{\mu{\mathcal M}}{c^2}+2\frac{\mu m}{(b-c)^2}+C\ ;
\end{equation}
}
\item{case 2: $0<z<b$:
\begin{equation}
0=\gamma_{\rm CC}(0,z)=-2\frac{m{\mathcal M}}{b^2}-2\frac{\mu{\mathcal M}}{c^2}+2\frac{\mu m}{(b-c)^2}+C\ ;
\end{equation}
}
\item{case 3: $b<z<c$:
\begin{equation}
0=\gamma_{\rm CC}(0,z)=2\frac{m{\mathcal M}}{b^2}-2\frac{\mu{\mathcal M}}{c^2}-2\frac{\mu m}{(b-c)^2}+C\ .
\end{equation}
}
\end{itemize}
The arbitrary constant $C$ cannot be uniquely chosen so that the function $\gamma_{\rm CCC}$ vanishes on the whole $z-$axis. 
In the following we choose $C=-[2m{\mathcal M}/b^2+2\mu{\mathcal M}/c^2+2\mu m/(b-c)^2]$, that makes it zero on the portions of axis outside the sources ($z>c$ and $z<0$).


\begin{figure}
\typeout{*** EPS figure 8}
\begin{center}
$\begin{array}{c@{\hspace{1in}}c}
\includegraphics[scale=0.3]{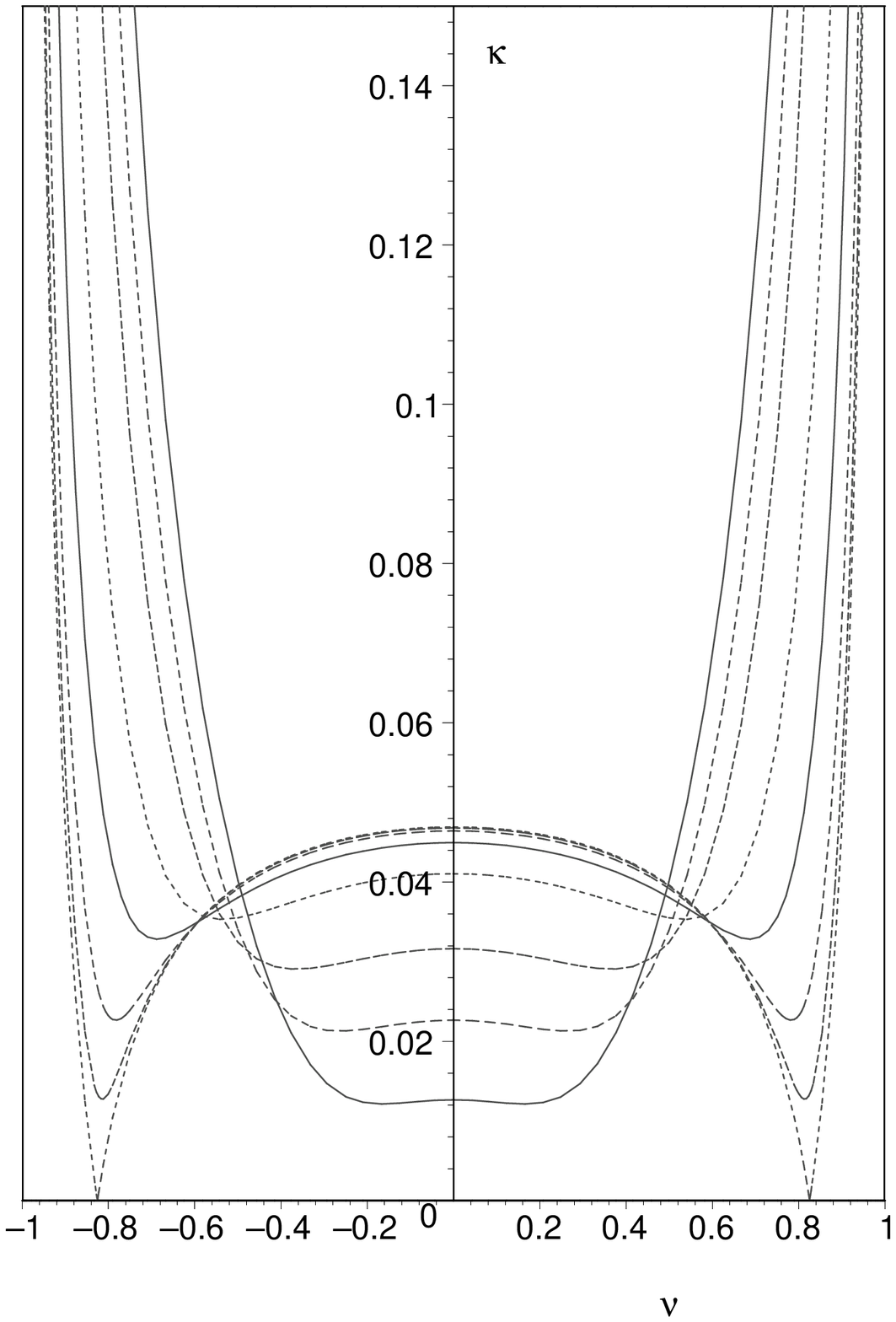}&
\includegraphics[scale=0.3]{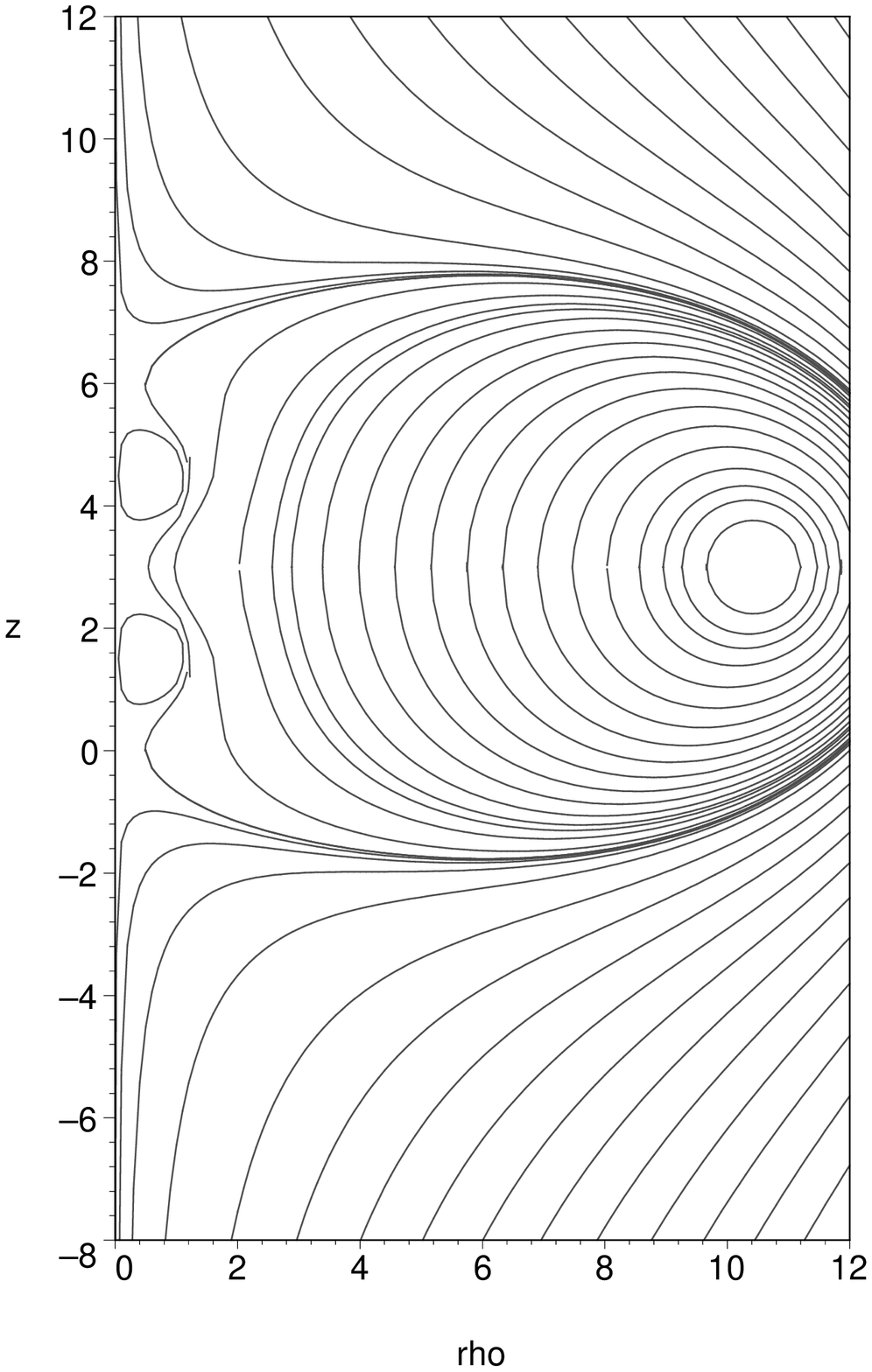}\\[0.4cm]
\mbox{(a)} & \mbox{(b)}
\end{array}$\\
\end{center}
\caption{
In Fig. (a) the magnitude of the acceleration in the case of background field generated by three Chazy-Curzon particles is shown as a function of $\nu$, namely for different orbits, corresponding to fixed values of $\rho=6$ and $z=[-10,-5.5, -3, -0.5, 1, 2, 2.5, 3]$ 
(as before, the various curves are ordered from bottom to top so that the one with $z=3$ is tangent to the $\kappa=0$ axis)
with  parameters $b=3$, $c=6$ for ${\mathcal M}=m=\mu=1$.
In Fig. (b) the lines of force of the acceleration field for the orbit corresponding to the value $\nu=0.6$
with  parameters $b=3$, $c=6$ for ${\mathcal M}=m=\mu=1$ are plotted.}  
\label{fig:8}
\end{figure}

The discussion of results in this case follows exactly that of the case of two Chazy-Curzon particles, and it is summarized here by Fig. \ref{fig:8}.
For instance, geodesics there exist in the case ${\mathcal M}=\mu$ and $c=2b$ on the middle plane $z=b$.

\section{Conclusions}

In this paper circular orbits in the gravitational background of two and three Chazy-Curzon particles placed at fixed positions along the $z-$axis are discussed.
Conditions for the existence of geodesics or other special orbits, as well as the analysis of the four acceleration through its lines of force are explicitly given. 
Of course this work can be easily generalized by considering even more Chazy-Curzon  particles or repeating the same analysis for the case of any number of Schwarzschild black holes or a mixed configuration of Schwarzschild black holes and Chazy-Curzon  particles. A systematic analysis of all these cases does not show important difference with respect to the case explored here.

\appendix
\section{Superposition of Schwarzschild black holes and Chazy-Curzon particles}

We give here the essential relations needed to repeat the discussion presented above for two or three Curzon particles for the case in which one has two Schwarschild black holes or one Schwarzschild black hole and one Curzon particle. The procedure can be easily generalized to take into account the case of three or more black holes or Curzon particles. 
In the case of two Chazy-Curzon particles the component $a^{\hat z}$ vanishes only for 
$z=b/2$ and for sources of equal masses $m={\mathcal M}$. It is easy to show that this fact remains true also for the case of two Schwarzschild black holes, but it is not more valid in the mixed situation of 
a Schwarzschild black hole and a Chazy-Curzon particle. 

\subsection{Superposition of two Schwarzschild black holes}

The solution corresponding to a linear superposition of two Schwarzschild black hole with masses ${\mathcal M}$ and $m$ and positions $z=0$ and $z=b$ on the $z-$axis respectively is given by metric (\ref{weylmetric}) with functions
\begin{eqnarray}
\label{psigammaSSb}
\psi=\psi_{\rm S}+\psi_{\rm S_b}\ , \qquad \gamma=\gamma_{\rm S}+\gamma_{\rm S_b}+\gamma_{\rm SS_b}\ ,
\end{eqnarray}
where  
\begin{eqnarray}
\label{SSbsol}
\psi_{\rm S}&=&\frac12\ln{\left[\frac{R_1^{+}+R_1^{-}-2{\mathcal M}}{R_1^{+}+R_1^{-}+2{\mathcal M}}\right]}\ , \qquad
\gamma_{\rm S}=\frac12\ln{\left[\frac{(R_1^{+}+R_1^{-})^2-4{\mathcal M}^2}{4R_1^{+}R_1^{-}}\right]}\nonumber\\
\psi_{\rm S_b}&=&\frac12\ln{\left[\frac{R_2^{+}+R_2^{-}-2m}{R_2^{+}+R_2^{-}+2m}\right]}\ , \qquad
\gamma_{\rm S_b}=\frac12\ln{\left[\frac{(R_2^{+}+R_2^{-})^2-4m^2}{4R_2^{+}R_2^{-}}\right]}\nonumber\\
\gamma_{\rm SS_b}&=&\frac12\ln{\left[\frac{E_{(1^{+},2^{-})}E_{(1^{-},2^{+})}}{E_{(1^{+},2^{+})}E_{(1^{-},2^{-})}}\right]}+C \ , \qquad
E_{(1^{\pm},2^{\pm})}=\rho^2+R_1^{\pm}R_2^{\pm}+Z_1^{\pm}Z_2^{\pm}\nonumber\\ 
R_1^{\pm}&=&\sqrt{\rho^2+(Z_1^{\pm})^2}\ , \qquad
R_2^{\pm}=\sqrt{\rho^2+(Z_2^{\pm})^2}\nonumber\\
Z_1^{\pm}&=&z\pm{\mathcal M}\ , \qquad
Z_2^{\pm}=z-(b\mp m)\ ,
\end{eqnarray}
the function $\gamma_{\rm SS_b}$ being obtained by solving Einstein's equations (\ref{einsteqs}).
The value of arbitrary constant $C$ can be determined by imposing the regularity condition (\ref{regcond}), from which we have
\begin{equation}
0=\lim_{\rho\rightarrow0}\gamma_{\rm SS_b}=\frac12\ln{\left[\frac{E_{(1^{+},2^{-})}(0,z)E_{(1^{-},2^{+})}(0,z)}{E_{(1^{+},2^{+})}(0,z)E_{(1^{-},2^{-})}(0,z)}\right]}+C\ ,
\end{equation}
with
\begin{equation}
E_{(1^{\pm},2^{\pm})}(0,z)=|Z_1^{\pm}||Z_2^{\pm}|+Z_1^{\pm}Z_2^{\pm}\ .
\end{equation}
The relevant cases are the following:
\begin{itemize}
\item{case 1: $z>b+m$ (and, analogously, $z<-{\mathcal M}$):
\begin{equation}
0=\gamma_{\rm SS_b}(0,z)=C\ ;
\end{equation}
}
\item{case 2: ${\mathcal M}<z<b-m$:
\begin{equation}
0=\gamma_{\rm SS_b}(0,z)=\ln{\left[\frac{b^2-({\mathcal M}+m)^2}{b^2-({\mathcal M}-m)^2}\right]}+C\ .
\end{equation}
}
\end{itemize}
A unique choice of the arbitrary constant $C$ allowing to make zero the function $\gamma_{\rm SS_b}$ on the whole $z$-axis does not exist: in fact, it vanishes on the segment ${\mathcal M}<z<b-m$ between the sources for $C=-\ln{([b^2-({\mathcal M}+m)^2]/[b^2-({\mathcal M}-m)^2])}$, and outside them (that is, for $z>b+m$ and $z<-{\mathcal M}$) if $C$ is chosen to be equal to zero.

\subsection{Superposition of a Schwarzschild black hole and a Chazy-Curzon particle}

The solution corresponding to a point particle of mass $m$ suspended on the symmetry axis at $z=b$ above the horizon of a Schwarzschild black hole with mass ${\mathcal M}$ at the origin is given by metric (\ref{weylmetric}) with functions
\begin{eqnarray}
\label{psigammaSCb}
\psi=\psi_{\rm S}+\psi_{\rm C_b}\ , \qquad \gamma=\gamma_{\rm S}+\gamma_{\rm C_b}+\gamma_{\rm SC_b}\ ,
\end{eqnarray}
where $\psi_{\rm C_b}$, $\gamma_{\rm C_b}$, and $\psi_{\rm S}$, $\gamma_{\rm S}$ are given by (\ref{ccsolw}) and (\ref{SSbsol}) respectively, while $\gamma_{\rm SC_b}$ can be obtained by solving Einstein's equations (\ref{einsteqs}):
\begin{equation}
\gamma_{\rm SC_b}=-\frac{m}{b^2-{\mathcal M}^2}\frac{(b+{\mathcal M})R_1^{-}-(b-{\mathcal M})R_1^{+}}{R_{\rm C_b}}+C\ .
\end{equation}
The value of arbitrary constant $C$ can be determined by imposing the regularity condition (\ref{regcond}),
from which we have
\begin{equation}
0=\lim_{\rho\rightarrow0}\gamma_{\rm SC_b}=-\frac{m}{b^2-{\mathcal M}^2}\frac{|Z_1^{-}|(b+{\mathcal M})-|Z_1^{+}|(b-{\mathcal M})}{|z-b|}+C\ .
\end{equation}
The relevant cases are the following ones:
\begin{itemize}
\item{case 1: $z>b$ (and, analogously, $z<-{\mathcal M}$):
\begin{equation}
0=\gamma_{\rm SC_b}(0,z)=-2\frac{m{\mathcal M}}{b^2-{\mathcal M}^2}+C\ ;
\end{equation}
}
\item{case 2: ${\mathcal M}<z<b$:
\begin{equation}
0=\gamma_{\rm SC_b}(0,z)=2\frac{m{\mathcal M}}{b^2-{\mathcal M}^2}+C\ .
\end{equation}
}
\end{itemize}
We can make $\gamma_{\rm SC_b}$ zero between the mass and the Schwarzschild source by choosing $C=-2m{\mathcal M}/(b^2-{\mathcal M}^2)$, or outside them by taking $C$ to have the opposite sign, but $C$ can not be chosen so that $\gamma_{\rm SC_b}$ vanishes on the whole $z$-axis.


\begin{thebibliography}{00}

\bibitem{weyl}
Weyl H., \textit{Ann.\ Phys.,\ Lpz.} {\bf 54}, 117 (1918).

\bibitem{exactsols}
Stephani H., Kramer D., McCallum M.A.H., Hoenselaers C. and Hertl E.,  
\textit{Exact solutions of Einstein's field equations},
Cambridge University Press, Cambridge (1979).

\bibitem{letelier}
Letelier P.S. and Oliveira S.R.,
{\it Class.\ Quantum\ Grav.\/}, {\bf 15}, 421 (1998).

\bibitem{sokolov}
Sokolov D.D. and Starobinskii A.A.,
{\it Sov.\ Phys.\ Dokl.\/}, {\bf 22}, 312 (1977).

\bibitem{israel}
Israel W.,  
{\it Phys.\ Rev.\/}, {\bf D15}, 935 (1977).

\bibitem{sem1}
Semer\'ak O., Zellerin T. and ${\check {\rm Z}}$\'a${\check {\rm c}}$ek  M.,
{\it MNRAS} {\bf 308}, 691 (1999).

\bibitem{sem2}
Semer\'ak O., ${\check {\rm Z}}$\'a${\check {\rm c}}$ek M. and Zellerin T.,
{\it MNRAS} {\bf 308}, 705 (1999).

\bibitem{iyevis}
Iyer B.R.  and Vishveshwara C.V.,  
{\it Phys.\ Rev.\/}, {\bf D48}, 5706 (1993).

\bibitem{def95}
de Felice F., 
{\it Class.\ Quantum Grav.\/}, {\bf 12}, 1119 (1991);
de Felice F. and Usseglio-Tomasset S.,
{\it Class.\ Quantum Grav.\/}, {\bf 8}, 1871 (1991).

\bibitem{pag98}
Page D.,
{\it Class.\ Quantum\ Grav.\/}, {\bf 15}, 1669 (1998).

\bibitem{sem98}
Semer\'ak O.,
{\it Gen.\ Relativ.\ Grav.\/}, {\bf 30}, 1203 (1998).

\bibitem{bjdf0}
Bini D., de Felice F. and Jantzen R.T., 
{\it Centripetal acceleration and centrifugal force in general relativity}
in {\it Nonlinear Gravitodynamics. The Lense-Thirring effect\/}, 
Ed. Ruffini R. and Sigismondi C. (Singapore: World Scientific), 2003.

\bibitem{circfs}
Bini D., Jantzen R.T. and Merloni A.,
{\it Class.\ Quantum Grav.\/}, {\bf 16}, 1333 (1999).

\bibitem{bjdf}
Bini D., de Felice F. and Jantzen R.T., 
{\it Class.\ Quantum Grav.\/}, {\bf 16}, 2105 (1999).

\bibitem{mfg}
Jantzen R.T., Carini P. and Bini D.,
{\it Ann.\ Phys.\ (N.Y.)\/}, {\bf 215}, 1 (1992).

\bibitem{idcf1}
Bini D., Carini P. and Jantzen R.T.,
{\it Int.\ J.\ Mod.\ Phys.\/}, {\bf D6}, 1 (1997).

\bibitem{idcf2}
Bini D., Carini P. and Jantzen R.T.,
{\it Int.\ J.\ Mod.\ Phys.\/}, {\bf D6}, 143 (1997).

\bibitem{chazy}
Chazy M.,
{\it Bull.\ Soc.\ Math.\ France\/}, {\bf 52}, 17 (1924).

\bibitem{curzon}
Curzon H.,
{\it Proc.\ London\ Math.\ Soc.\/}, {\bf 23}, 477 (1924).

\bibitem{scott}
Scott S.M. and Szekeres P.,
{\it Gen. Relativ. Grav.}, {\bf 18}, 557 (1986);
{\it Gen. Relativ. Grav.}, {\bf 18}, 571 (1986).

\bibitem{abram}
Abramowicz M.A., Carter B. and Lasota J.P.,
{\it Gen. Relativ. Grav.}, {\bf 20}, 1173 (1988).

\bibitem{karasetal}
Karas V., Hur\'e J.M. and Semer\'ak O.,
{\it Gravitating discs around black holes},
e-print gr-qc/0401345


\end{thebibliography}
\end{document}